\documentclass[letterpaper,11pt]{article}
\usepackage{caption}
\pdfoutput=1 
\usepackage{xcolor}
\usepackage{enumerate}
\usepackage{framed}
\usepackage{mdframed}
\usepackage{amsmath}
\usepackage{amsfonts}
\usepackage{mathrsfs}
\usepackage{amssymb}
\usepackage[normalem]{ulem}
\usepackage{graphicx, rotating}
\usepackage{epsfig}
\usepackage{latexsym}
\usepackage{graphicx}
\usepackage{color}
\usepackage{amsmath,bm,amssymb}
\usepackage{cite}
\usepackage{slashed}
\usepackage{hyperref}
\usepackage{epstopdf}
\usepackage[title]{appendix}
\usepackage[font=footnotesize,labelfont=]{caption}
\AppendGraphicsExtensions{.tif}
\hypersetup{colorlinks=true, citecolor=bluscuro, linkcolor=black, urlcolor=bluscuro}
\definecolor{rossos}{cmyk}{0,1,1,0.55}
\definecolor{bluscuro}{rgb}{0.15, 0.2, .85}
\definecolor{bluchiaro}{cmyk}{1,.3,0.,0.1}
\definecolor{Green}{rgb}{0, 0.65, 0.31}

\newcommand{\be}{\begin{equation}}
\newcommand{\ee}{\end{equation}}
\newcommand{\bea}{\begin{eqnarray}}
\newcommand{\eea}{\end{eqnarray}}
\newcommand{\beas}{\begin{eqnarray*}}
\newcommand{\eeas}{\end{eqnarray*}}

\newcommand{\rd}{{\rm d}}

\newcommand{\lp}{\left (}
\newcommand{\rp}{\right )}

\begin{document}
\def\thefootnote{\fnsymbol{footnote}}

\begin{center}
\LARGE{\textbf{Non-linearities in the tidal Love numbers of black holes}} \\[0.5cm]
 
\large{Valerio De Luca\footnote{vdeluca@sas.upenn.edu}, Justin Khoury\footnote{jkhoury@sas.upenn.edu} and Sam S. C. Wong\footnote{scswong@sas.upenn.edu}}
\\[0.5cm]

\small{
\textit{Center for Particle Cosmology, Department of Physics and Astronomy, University of Pennsylvania,\\ Philadelphia, PA 19104}}

\vspace{.2cm}

\end{center}

\vspace{.6cm}

\hrule \vspace{0.2cm}
\centerline{\small{\bf Abstract}}
{\small\noindent 
Tidal Love numbers describe the linear response of a compact object under the presence of external tidal perturbations, and they are found to vanish exactly for black holes within General Relativity. In this paper we investigate the tidal deformability of neutral black holes when non-linearities in the theory are taken into account. As a case in point, we consider scalar tidal perturbations on the black hole background, and find that the tidal Love numbers may be non vanishing depending on the scalar interactions in the bulk theory. Remarkably, for non-linear sigma models, we find that the tidal Love numbers vanish to all orders in perturbation theory.
}

\vspace{0.3cm}
\noindent
\hrule
\def\thefootnote{\arabic{footnote}}
\setcounter{footnote}{0}

\section{Introduction}
Black holes and gravitational waves represent two of the most striking predictions of General Relativity, and they come along when one considers a binary system of black holes orbiting around each other and emitting gravity waves as the coalescence proceeds. The measurement of this gravitational radiation therefore provides a powerful tool to shed light on the nature of gravity in its most extreme regime and on the search for signatures of new physics~\cite{LIGOScientific:2021sio}. 

Within this theory, the black hole mass, angular momentum, and electric charge uniquely define their multipolar structure. The latter is not relevant during the early stages of the inspiral phase of a compact binary, since the two bodies behave as point masses~\cite{PhysRev.136.B1224}.
However, as the orbital separation sufficiently decreases due to gravitational wave emission, the tidal interactions between the two bodies become important, and higher-order post-Newtonian corrections come into action. These tidal effects are usually described in terms of the tidal Love numbers~\cite{poisson_will_2014}. The tidal Love numbers depend on the internal properties and structure of the deformed compact object, and are found to impact gravitational wave emission at fifth post-Newtonian order~\cite{Flanagan:2007ix}. 

Assuming General Relativity, the tidal Love numbers of non-rotating and spinning black holes are found to be exactly zero~\cite{Binnington:2009bb,Damour:2009vw,Damour:2009va,Pani:2015hfa,Pani:2015nua,Gurlebeck:2015xpa,Porto:2016zng,LeTiec:2020spy, Chia:2020yla,LeTiec:2020bos}. This result has generated a problem of “naturalness” in the gravitational theory~\cite{Porto:2016zng} and  reveals underlying hidden symmetries of General Relativity~\cite{Hui:2020xxx,Charalambous:2021mea,Charalambous:2021kcz,Hui:2021vcv,Hui:2022vbh,Charalambous:2022rre,Ivanov:2022qqt,Katagiri:2022vyz, Bonelli:2021uvf,Kehagias:2022ndy,BenAchour:2022uqo,Berens:2022ebl}.
This property is, however, fragile, since it is broken in higher dimensions~\cite{Kol:2011vg,Cardoso:2019vof, Hui:2020xxx,Pereniguez:2021xcj,Charalambous:2023jgq,Rodriguez:2023xjd}, in the context of modified gravity~\cite{Cardoso:2017cfl,Cardoso:2018ptl,Cvetic:2021vxa,DeLuca:2022tkm}, or due to environmental effects around the black holes~\cite{Baumann:2018vus,Cardoso:2019upw,DeLuca:2021ite,DeLuca:2022xlz,Katagiri:2023yzm}.

Known results on tidal Love numbers usually assume linear response theory, based on a weak external tidal field, and consider free massless perturbations on the black hole background. General Relativity, on the other hand, is a non-linear theory.
The issue of non-linearities has been deeply investigated in the context of black hole perturbation theory and quasi-normal modes~\cite{1967PThPh..37..831T,Gleiser:1996yc,Gleiser:1998rw,Campanelli:1998jv,Nakano:2007cj,Brizuela:2007zza,Lousto:2008vw,Loutrel:2020wbw,Ripley:2020xby}. It has recently received further interest since black hole merger simulations have shown that not only first-order but also second-order effects are relevant to describe ringdowns~\cite{Mitman:2022qdl,Cheung:2022rbm,Ma:2022wpv,Lagos:2022otp,Kehagias:2023ctr,Baibhav:2023clw,Kehagias:2023mcl}.

In this paper we investigate the role of non-linearities in the context of tidal deformations. For simplicity, we focus on self-interacting scalar fields on a fixed Schwarzschild background, treating interactions perturbatively. Scalars are simpler to deal with, since they avoid the issue of parity-even and -odd perturbation mixing that would arise for spin-1 (electromagnetic) and spin-2 (gravitational) non-linearities, as well as issues of gauge invariance. 

In a linear theory, a basis of two linearly independent solutions can be chosen to have the correct growing/decaying behavior at spatial infinity and regularity/divergence behavior at some finite distance. They correspond to the usual ``external source" and ``response" components. The Love number is easily defined as the relative coefficients of these two independent solutions. However, one can no longer separate them in a non-linear theory, such that defining the Love number becomes subtle. Despite this difficulty, the presence of fall-off tails may still signal a tidal response in a non-linear way. Therefore, for different self-interactions, we compute the solution of the scalar perturbations on the black hole background and determine their asymptotic behavior to look for tails at each multipole.

We consider three families of interacting scalar theories:

\begin{itemize}

    \item {\bf Potential interactions}. Specifically, we consider power-law operators~$\phi^n$. In this case, we find that a tail is generated at second-order in perturbation theory.

    \item {\bf Derivative interactions}. Specifically, we consider~$(\partial\phi)^4$ as the simplest shift-symmetric interaction. In this case, we find that a tail is generated at first order in perturbation theory. This is an immediate consequence of the higher-derivative nature of the operator.

    \item {\bf Non-linear sigma model},~$G_{IJ} (\phi) \partial\phi^I \partial\phi^J$. We place no restriction on the target space metric~$G_{IJ}(\phi)$, other than it is non-singular in the field region of interest, such that we can adopt Riemann normal coordinates. In particular,~$G_{IJ}(\phi)$ need not be flat. Remarkably, in this case {\it we will prove the absence of a tail to all orders in perturbation theory.}
    
\end{itemize}

The non-linear sigma model is the closest scalar analogue to General Relativity, in the sense that, exactly like the Einstein-Hilbert term,~$G_{IJ} (\phi) \partial\phi^I \partial\phi^J$ includes exactly two derivatives and possibly all powers of the field. The fact that we find no tail in this case is striking and suggestive of an underlying symmetry.

The paper is organized as follows. In Sec.~\ref{review sec} we briefly review the notion of Love numbers, both in Newtonian gravity and in General Relativity. In Sec.~\ref{form sec} we provide details about the non-linearities involved in the computation of the Love numbers for a scalar tidal perturbation. We study a few illustrative examples of non-linear scalar theories in Sec.~\ref{examples}, while dedicating Sec.~\ref{NLSM sec} to the study of the non-linear sigma model. Finally, we conclude in Sec.~\ref{conclu sec}. Three Appendices are devoted to technical details.

\section{Brief review of tidal Love numbers}
\label{review sec}

In this Section we review the formalism to compute the tidal Love numbers for a black hole. We start by discussing their analogue with the electric polarizability of a material, and then define the Love numbers both in the Newtonian and full relativistic regime. For the latter, we focus on massless spin-0 tidal perturbations, assuming for simplicity a Schwarzschild black hole background. The interested reader can find a more comprehensive discussion in Refs.~\cite{Hui:2020xxx, Charalambous:2021mea,Charalambous:2022rre}.

\subsection{Electromagnetic response}

Computing Love numbers is in fact closely similar to the calculation of the electric polarizability of a material, such as a dielectric material under a static external electric field. The static problem outside the material (in vacuum) is simply given by Laplace's equation for the electric potential~$\Phi$, 
\be
\nabla^2 \Phi = 0\,.
\ee
The general solution is given by 
\begin{align} \label{eqn:laplacesol}
    \Phi = \sum_{\ell,  m} c_{\ell m}\left( r^\ell + \frac{\lambda_{\ell m}}{r^{\ell +1}}  \right) Y_{\ell m}(\theta, \varphi)\,, 
\end{align}
where~$Y_{\ell m}$'s are the real spherical harmonics. The first term~$c_{\ell m} r^\ell Y_{\ell m}$ is simply the potential for the external electric field. The second term~$c_{\ell m}\frac{\lambda_{\ell m}}{r^{\ell +1}}  Y_{\ell m}$ characterizes the induced multiple of the material, with~$\lambda_{\ell m}$ identified as the electric polarizability of the object. 

In practice, in order to calculate~$\lambda_{\ell m}$, one must specify the properties of the material. For instance, given a relation between the polarization vector and the electric field inside the material,
\begin{align}
    P^i = \chi^{(1)\,i}_{\quad\; j} E^{j} + \chi^{(2)\,i}_{\quad\; j k} E^{j}E^{k}+ \dots \,,
\end{align}
with~$\chi^{(n)}$ being the~$n$-th order susceptibility tensor, one can solve the electrostatic problem inside the material, and then match the solution with Eq.~\eqref{eqn:laplacesol} at the surface of the object to satisfy the  boundary condition of regularity. A well known example is a linear dielectric (${\vec P} = \chi {\vec E}$) solid sphere of radius~$R$ in an arbitrary external electric field. In this case,~$\Phi$ is solved by 
\begin{align}
    \Phi_{\rm outside} &= \sum_{\ell, \, m} V_{\ell m} \left( r^\ell - \frac{\chi \ell R^\ell}{ 2\ell +1 + \ell \chi   } \left(\frac{R}{r}\right)^{\ell +1} \right) Y_{\ell m}( \theta, \varphi) \,; \nonumber \\
     \Phi_{\rm inside} &= \sum_{\ell, \, m} V_{\ell m} \frac{2\ell +1}{2\ell +1 + \ell \chi }r^\ell Y_{\ell m}( \theta, \varphi) \,,
\end{align}
where~$V_{\ell m}$ quantifies the potential for the external electric field. It is clear that the induced multiple moment is proportional to the susceptibility~$\chi$. The strength of the induced multiple moment can be thought of as the electric counterpart of the gravitational Love number.

In the following we solve the analogous problem of a gravitational perturbation deforming a compact object, which we will assume to be a Schwarzschild black hole, both in Newtonian and Einstein gravity.

\subsection{Gravitational response: Newtonian limit}

The gravitational response of a spherically-symmetric body to external tidal perturbations is captured by the tidal Love numbers. Consider a spherical body of mass~$M$, possibly spinning, and assumed to be at the origin of a Cartesian coordinate frame. The body is subjected to an external gravitational field~$U_\text{\tiny ext}$ applied adiabatically. 

Assuming spherical symmetry, the field can be expanded in  multipole moments as
\be
U_\text{\tiny ext} =-\sum_{\ell = 2}^\infty \frac{(\ell-2)!}{\ell!}\mathcal{E}_{L} r^L\,,
\ee
in terms of its distance~$r$ from the body, the multi-index~$L\equiv i_1\cdots i_\ell$, and the symmetric trace-free multipole moments~$\mathcal{E}_{L}$. This external gravitational perturbation induces a deformation in the body, which develops internal multipole moments given by
\be
G I_{L} =\int {\rm d}^3x~\delta\rho(\vec x) x^{\langle L \rangle} \,,
\ee
where~$\delta\rho$ is the body's mass density perturbation, and~$x^{\langle L \rangle} \equiv x^{i_1}\cdots x^{i_\ell}$.

Expanding the external source and induced response in spherical harmonics~$Y_{\ell m}$,
\be
\mathcal{E}_{\ell m}\equiv \mathcal{E}_{L}\int_{\mathbb{S}^2} {\rm d} \Omega~n^L Y^{*}_{\ell m}\,;\quad 
I_{\ell m}\equiv I_{L}\int_{\mathbb{S}^2} {\rm d}\Omega~n^L Y^{*}_{\ell m}\,,
\ee
where~${\rm d} \Omega \equiv \sin\theta\,{\rm d}\theta\, {\rm d}\phi$, and~$n^i \equiv x^i/|\vec x|$, the total potential of the system can be written as
\be 
U_\text{\tiny tot}=
-\frac{G M}{r}
-\sum_{\ell=2}^\infty \sum_{m=-\ell}^\ell Y_{\ell m} 
\left[\frac{(\ell-2)!}{\ell !} \mathcal{E}_{\ell m}  r^\ell -
\frac{(2\ell-1)!!}{\ell !} 
\frac{G I_{\ell m}}{r^{\ell+1}}
 \right]\,.
\ee
Assuming that the external tidal perturbation is adiabatic and weak, linear response theory dictates that the response multipoles are proportional to the perturbing multipole moments as
\be
G I_{\ell m}\left(\omega\right)
=-\frac{\left(\ell-2\right)!}{(2\ell-1)!!}\lambda_{\ell m}(\omega) r_h^{2\ell+1} \mathcal{E}_{\ell m}\left(\omega\right)\,,
\ee
where~$\omega$ is the perturbation frequency. The size of the object,~$r_h$, will later be identified with the Schwarzschild radius in the black hole case, $r_h = 2 G M$.
The dimensionless coefficients~$\lambda_{\ell m}$ describe the tidal response and can be expressed in terms of~$\omega$ as
\be
\label{eq:klm}
	\lambda_{\ell m} \simeq k_{\ell m} + {\rm i}\nu_{\ell m}\left(\omega - m \Omega\right) + \dots\,,
\ee
where~$m$ is the azimuthal harmonic number, and~$\Omega$ the body's angular velocity. The real term in Eq.~\eqref{eq:klm} encodes the static response, and the corresponding coefficients~$k_{\ell m}$ are called tidal Love numbers. The imaginary contribution~${\rm i}\nu_{\ell m}$ describes dissipation effects. For the sake of our discussion we will assume static perturbations~$(\omega = 0)$ and therefore neglect dissipative effects.

The Newtonian regime considered so far represents the long-distance approximation to full General Relativity. In the following we will therefore discuss tidal Love numbers for a Schwarzschild black hole assuming a massless spin-0 perturbation in a full relativistic theory. 

\subsection{Gravitational response: General relativity}

We now turn to the gravitational response of a black hole in General Relativity, which for concreteness we assume to be Schwarzschild:
\be
{\rm d} s^2 = - f (r) {\rm d} t^2 + \frac{1}{f (r)} {\rm d}r^2 + r^2 {\rm d} \Omega^2\,; \qquad f(r) =  1-\frac{r_h}{r}\,.
\ee
Using black hole perturbation theory, one can consider massless fields perturbing the black hole geometry. The simplest example is a tidal spin-2 perturbation, which describes the presence of a companion in a binary system. 

For the sake of our discussion we will focus on a scalar tidal field~$\phi$, which is responsible for the generation of a scalar Love number.
The dynamics of a free real scalar field is governed by the action
\be
S = -\frac{1}{2}\int \rd^4 x \sqrt{-g} \,(\partial\phi)^2 \,.
\label{free scalar action}
\ee
Since the background is invariant under rotations, we can decompose~$\phi$ into spherical harmonics,
\be
\phi(x) = \sum_{\ell,m} \Psi(t,r)r^{-1}Y_{\ell m}(\theta,\varphi)\,.
\ee
The action can then be simplified by integrating over the angular variables. Introducing the tortoise coordinate~$r_\star$, with~$\rd r_\star = \rd r/f$, we obtain
\be
S = \sum_{\ell, m}\int\rd t\rd r_\star\left(\frac{1}{2} |\dot\Psi|^2 -\frac{1}{2}\left|\frac{\partial\Psi}{\partial r_\star}\right|^2 -\frac{1}{2}V_0(r)|\Psi|^2
\right)\,,
\ee
in terms of the scalar potential
\be
V_0(r) \equiv f\frac{\ell(\ell+1)}{r^2} +\frac{ff'}{r}\,,
\ee
where primes denote radial derivatives ($f' = {\rm d}f/{\rm d}r$). The resulting equation of motion takes the form of a Schr\"odinger equation
\be
\frac{\rd^2\Psi(r_\star)}{\rd r_\star^2} + \Big(\omega^2 - V_0(r)\Big)\Psi(r_\star)= 0\,.
\ee
In the zero-frequency limit, imposing regularity of the solution at the black hole horizon~$r_h$, and matching to the external source at spatial infinity~$r \to \infty$, one finds that~$\Psi$ can be expanded asymptotically as
\be
\label{scalarTLN}
\Psi(r) \simeq c_1 r^{\ell+1}\left[1+\cdots+ k_\text{\tiny S}^{(\ell)} \left(\frac{r}{r_h}\right)^{-2\ell-1}+\cdots\right]\,.
\ee
The first term~$\sim r^{\ell+1}$ denotes the external tidal field applied at spatial infinity, while the second term~$\sim r^{-\ell}$ encodes the response. The coefficient~$k_\text{\tiny S}^{(\ell)}$ denotes the scalar Love number. Its value depends on the assumed background geometry. For a Schwarzschild black hole~$k_\text{\tiny S}^{(\ell)}$ is found to vanish identically.

The procedure outlined above is called Newtonian matching, since the extraction of the Love numbers for the full relativistic case is based on the comparison between this series expansion and the non-relativistic gravitational potential. There is, however, an intrinsic ambiguity in this procedure, due to the overlap between the source series and the response contribution~\cite{Kol:2011vg,LeTiec:2020spy,Charalambous:2021mea}, where subleading corrections to the source appear to have the same power in~$r$ as the response in the physical case~$\ell \in \mathbb{N}$. In order to properly define the Love numbers through a matching procedure, an interesting approach consists of performing an analytic continuation to the unphysical region~$\ell \in \mathbb{R}$~\cite{LeTiec:2020spy}, where the source and response series do not overlap. Obtaining such a solution can, however, be challenging, and depends on the theory at hand. Furthermore, this definition of tidal Love numbers may also be problematic due to gauge invariance. 

In order to get around these issues and provide a gauge invariant definition, one can instead identify the relevant operators describing tidal effects in the point-particle effective field theory approach. This approach is based on the realization that, at very large distance, a black hole behaves as a point particle, and corrections due to its finite size and internal structure are encoded in higher-derivative operators in the effective theory. 

The most general action describing the interactions between an external scalar field and the point particle worldline, up to field redefinitions and down to second order in the bulk scalar field, is given by~\cite{Hui:2020xxx}
\be
S = -\frac{1}{2}\int \rd^4 x \sqrt{-g} \,(\partial\phi)^2+ \int \rd \tau e \left[\frac{1}{2}e^{-2} \dot x^\mu\dot x_\mu  - \frac{m^2}{2} + g \phi +\sum_{\ell=1}^{\infty} \frac{\lambda_\ell}{2\ell!} \left(  \partial_{(a_1}\cdots \partial_{a_\ell)_T} \phi \right)^2\right]\,,
\label{eq:pointscalaraction}
\ee
where~$e$ denotes the vielbein, and~$(\cdots)_T$ the symmetrized traceless component of the enclosed indices. The bulk action is just that of the free scalar field, Eq.~\eqref{free scalar action}. The first two terms in the worldline action describe the worldline trajectory; the term~$g\phi$ encodes the scalar hair carried by the point particle which, in the black hole context, is absent due to the no-hair theorems~\cite{Bekenstein:1971hc,Bekenstein:1995un,Hui:2012qt}; and the last term is the leading-order finite-size effective operator, where the Wilson coefficients~$\lambda_\ell$ indicate the worldline definitions of black hole static response coefficients. From a Feynman diagrammatic level, one can interpret this quadratic operator as a vertex, with one of the external fields behaving as a background and the other as a response at infinity.

Solving the corresponding equation of motion in the zero-frequency limit one obtains the full field solution~$\phi$ as
\be
\phi(\vec x) = c_{a_1\cdots a_\ell}x^{a_1}\cdots x^{a_\ell}\left[1+\lambda_\ell(-1)^\ell\frac{2^{\ell-2}\Gamma(\tfrac{1}{2})^2}{\pi^{3/2}\Gamma(\tfrac{1}{2}-\ell)}\lvert\vec x\rvert^{-2\ell-1}\right]\,; \qquad x = r/r_h\,,
\label{eq:eftscalar}
\ee
where $c_{a_1\cdots a_\ell}$ is a symmetric traceless tensor.
This can be matched to the full relativistic solution obtained in Eq.~\eqref{scalarTLN} to extract the gauge-invariant scalar Love numbers as~\cite{Hui:2020xxx}
\be
\lambda_\ell = k_\text{\tiny S}^{(\ell)} (-1)^\ell \frac{\pi^{3/2}}{2^{\ell-2}} \frac{\Gamma(\frac{1}{2}-\ell)}{ \Gamma(\frac{1}{2})^2} r_h^{2\ell+1}\,.
\ee
This result holds at the linear level. In principle, by adding terms with more powers of the fields to the effective action, one can study the non-linear response of the system. This is the approach we will pursue in the next Section.

\section{Non-linearities in the tidal Love numbers: Formalism}
\label{form sec}

In this Section we describe two different kinds of non-linearities that may affect the computation of the tidal Love numbers. The first kind of non-linearity is related to the assumption of linear response, and arises when one goes to the next-to-leading order in the applied external field. The second kind is instead based on a theory which is inherently non-linear, such as General Relativity, but keeping the assumption of linear response theory. 

These non-linearities are described by two different expansion parameters: for the first kind, we are perturbatively expanding in the tidal response, which is proportional to the size or compactness of the tidally deformed object; for the second kind, we are  expanding in the strength of the external applied field, which is parametrized by the coupling constant of the theory at hand, {\it e.g.}, the Planck mass for General Relativity. This emphasizes the different nature of the non-linearities we will describe in this Section.

\subsection{Beyond linear response theory}

One of the main assumptions in the computation of the Love numbers relies on linear response theory, according to which the amount of deformation captured in the mass quadrupole is proportional to the strength of the external applied field.

The same assumption is usually performed also in electromagnetism, where the electric dipole at leading order is proportional to the external electric field. However, one can generalize the response by including higher order corrections in the electric field. Indeed, 
the non-linear response of a conducting body to an external electric field is described by the electric dipole, with a power expansion in the electric field as~\cite{10.5555/1817101}
\be
D_\mu = \lambda^{(1)}_{\mu \nu} (t, \vec x) E_{\nu} (t, \vec x) + \lambda^{(2)}_{\mu \nu \rho} (t, \vec x) E_{\nu} (t, \vec x) E_{\rho} (t, \vec x) + \lambda^{(3)}_{\mu \nu \rho \sigma} (t, \vec x) E_{\nu} (t, \vec x) E_{\rho} (t, \vec x) E_{\sigma} (t, \vec x) + \dots\,,
\ee
where~$E_{\mu} = F_{\mu\nu}u^\nu$ covariantly defines the electric field in the body's rest frame, and~$\lambda^{(n)}$
are the susceptibility tensors, which are symmetric in their indices. We shall refer to~$D_\mu^{(2)} = \lambda^{(2)}_{\mu \nu \rho} (t, \vec x) E_{\nu} (t, \vec x) E_{\rho} (t, \vec x)$ as the second-order non-linear dipole moment, to~$D_\mu^{(3)} =\lambda^{(3)}_{\mu \nu \rho \sigma} (t, \vec x) E_{\nu} (t, \vec x) E_{\rho} (t, \vec x) E_{\sigma} (t, \vec x)$ as the third-order non-linear dipole moment, and so on for higher-order terms.

Assuming Lorentz invariance and parity conservation, the index structure of the tensor susceptibilities is constrained, such that they can be decomposed in a set of scalar susceptibilities as~\cite{Bern:2020uwk}
\begin{align}
\lambda^{(1)}_{\mu \nu} & = \lambda^{(1)}_0 g_{\mu \nu} + \lambda^{(1)}_1 q_{1,\mu} q_{2,\nu} \,; \nonumber \\
\lambda^{(2)}_{\mu \nu \rho} & =  \lambda^{(2)}_0 \left( g_{\mu \nu} q_{3,\rho} + g_{\nu \rho} q_{1,\mu} + g_{\mu \rho} q_{2,\nu}  \right)\,; \nonumber \\
\lambda^{(3)}_{\mu \nu \rho \sigma} & = \lambda^{(3)}_0 g_{(\mu \nu} g_{\rho \sigma)} + \lambda^{(3)}_1 (g_{\mu \nu} q_{3,\rho} q_{4,\sigma} + {\rm perms}) + \lambda^{(3)}_2 q_{1,\mu} q_{2,\nu} q_{3,\rho} q_{4,\sigma}\,,
\end{align}
in terms of the background metric~$g_{\mu \nu}$ and body's momentum~$q^\mu$.

The analogy of the electric dipole in the context of gravity is provided by the quadrupole moment, which can be as well expanded in a power series 
\be
I_{\mu \nu} (t, \vec x) = \lambda^{(1)}_{\mu \nu \rho \sigma} (t, \vec x) E_{\rho \sigma} (t, \vec x) + \lambda^{(2)}_{\mu \nu \rho \sigma \kappa \delta} (t, \vec x) E_{\rho \sigma} (t, \vec x) E_{\kappa \delta} (t, \vec x) + \dots\,,
\ee
in terms of the electric and magnetic fields, built from the Weyl tensor~$C_{\mu \rho \nu \sigma}$ as~$E_{\mu \nu} = C_{\mu \rho \nu \sigma} u^\rho u^\sigma$ and~$B_{\mu \nu} = (*C)_{\mu \rho \nu \sigma} u^\rho u^\sigma = \frac{1}{2} \epsilon_{\alpha \beta \rho \mu} C^{\alpha \beta}_{\sigma \nu} u^\rho u^\sigma$, where~$u^\mu$ denotes the body four-velocity.

From this analogy one can appreciate the role of higher operators in determining the non-linear response of the object to the external perturbation.
Such operators also arise in an effective field theory approach, as higher-derivative operators in the Lagrangian. Considering a scalar field~$\phi$, the worldline action would receive next-to-leading terms built out of derivative of the scalar field as
\begin{align}
S &\supset \int {\rm d} \tau \, e \Bigg[ \sum_\ell \frac{\lambda_\ell^{(1)}}{2 \ell !} \left(\partial_{(a_1} \dots \partial_{a_{\ell)_T}} \phi\right)^2 \nonumber \\
& + \sum_{\ell_1,\ell_2,\ell_3} \frac{\lambda_{\ell_1\ell_2\ell_3}^{(2)}}{2 \ell_1 ! 2 \ell_2 ! 2 \ell_3 !} \left(\partial_{(a_1} \dots \partial_{a_{\ell_1)}} \phi \right) \left(\partial_{(b_1} \dots \partial_{b_{\ell_2)}} \phi\right) \left(\partial_{(c_1} \dots \partial_{c_{\ell_3)}} \phi\right)  \nonumber \\
& + \sum_{\ell_1,\ell_2,\ell_3, \ell_4} \frac{\lambda_{\ell_1\ell_2\ell_3\ell_4}^{(3)}}{2 \ell_1 ! 2 \ell_2 ! 2 \ell_3 ! 2 \ell_4 !} \left(\partial_{(a_1} \dots \partial_{a_{\ell_1)}} \phi\right) \left(\partial_{(b_1} \dots \partial_{b_{\ell_2)}} \phi\right) \left(\partial_{(c_1} \dots \partial_{c_{\ell_3)}} \phi\right)  
 \left(\partial_{(d_1} \dots \partial_{d_{\ell_4)}} \phi\right) 
 \nonumber \\
 & + \dots \Bigg]\,,
\end{align}
where the lower indices~$a_i,\, b_i,\, c_i$ and~$d_i$ are properly contracted in a Lorentz invariant and traceless way, such that~$\ell_3 = |\ell_1-\ell_2| ,\dots , \ell_1+\ell_2$ for the second line, and similarly for the third line. The list of coefficients~$\lambda_{\ell_1\ell_2\ell_3}^{(2)}$,~$\lambda_{\ell_1\ell_2\ell_3\ell_4}^{(3)}$,~$\dots$ capture the non-linear tidal Love numbers. From a Feynman diagrammatic level, we would expect in this case diagrams with more lines either as a background, {\it i.e.}, a non-linear source, or as a non-linear response.

In the following we will not consider this kind of non-linearity, which we expect to be subdominant compared to the one at the linear level, due to the higher number of derivatives involved. We leave their study to future work.

\subsection{Non-linear bulk theory}
\label{nonlin bulk}

The second kind of non-linearities may arise from the theory one considers. For example, when studying a black hole perturbed by an external spin-2 tidal field, such as its companion in a binary system, the theory at hand, {\it i.e.}, Einstein gravity, is inherently non-linear. This non-linearity has been investigated extensively in the literature in the context of black hole perturbation theory and quasi-normal modes, {\it e.g.},~\cite{1967PThPh..37..831T,Gleiser:1996yc,Gleiser:1998rw,Campanelli:1998jv,Nakano:2007cj,Brizuela:2007zza,Lousto:2008vw,Loutrel:2020wbw,Ripley:2020xby}. Although the dimensionless amplitude of the metric perturbations are negligibly small when we observe them at detectors, they are relatively large near the black hole, so non-linearities may play an important role at gravitational wave detectors, see in particular Refs.~\cite{Mitman:2022qdl,Cheung:2022rbm,Ma:2022wpv,Lagos:2022otp,Kehagias:2023ctr,Baibhav:2023clw,Kehagias:2023mcl} for recent developments in the context of black hole ringdown.

To illustrate this, let us focus on parity-odd perturbations. At the linear level the metric perturbation~$\Psi_\text{\tiny RW}^{(1)}$ around a Schwarzschild black hole satisfies the familiar Regge-Wheeler equation~\cite{Regge:1957td},
\be
\bigg(\frac{\partial^2}{\partial r_\star^2} - V_\text{\tiny RW}(r) \bigg)\Psi_\text{\tiny RW}^{(1)} = 0\,.
\ee
The Regge-Wheeler potential~$V_\text{\tiny RW}$ takes a form similar to the one obtained above for a scalar field. One can go beyond linear level, and focus on “higher-order” terms in the metric perturbations,
\be
\Psi_\text{\tiny RW} = \Psi_\text{\tiny RW}^{(1)} + \epsilon \Psi_\text{\tiny RW}^{(2)}\,.
\ee
From a perturbative expansion it was shown that the second order perturbations satisfy the same linear set of equations as the first order perturbations, but with additional terms quadratic in the first-order perturbations, that may be thought of as “sources” for these equations. Explicitly,~$\Psi_\text{\tiny RW}^{(2)}$ satisfies a Regge-Wheeler equation, 
\be
\bigg(\frac{\partial^2}{\partial r_\star^2} - V_\text{\tiny RW}(r) \bigg)\Psi_\text{\tiny RW}^{(2)} = S\left(\Psi_\text{\tiny RW}^{(1)\,2}\right)\,,
\ee
which now includes a non-linear source in the first order perturbations~$S(\Psi_\text{\tiny RW}^{(1)\,2})$.\footnote{In contrast with first order perturbations, at second order tensor modes are not gauge invariant. This requires the definition of gauge-invariant quantities in order to make gravitational wave predictions~\cite{Gleiser:1998rw, Campanelli:1998jv,Brizuela:2007zza,Nakano:2007cj,Loutrel:2020wbw,Kehagias:2023mcl}.} 

In the rest of our paper we focus on this second kind of non-linearity in the context of interacting scalar tidal perturbations around a Schwarzschild black hole. The corresponding procedure for tensor perturbations is left to future work. 

Allowing for bulk self-interactions characterized by an operator~$\mathcal{O}(\phi)$, the worldline effective field theory action~\eqref{eq:pointscalaraction} generalizes to
\begin{align}
S &= -\frac{1}{2}\int {\rm d}^4 x \sqrt{-g} \Big[ \,(\partial\phi)^2 + \alpha \, \mathcal{O}(\phi) \Big] \nonumber\\
& + \int {\rm d} \tau e \left[\frac{1}{2}e^{-2} \dot x^\mu\dot x_\mu - \frac{m^2}{2} + g\phi + \sum_{\ell=1}^{\infty} \frac{\lambda_\ell}{2\ell!} \left(  \partial_{(a_1}\cdots \partial_{a_\ell)_T} \phi \right)^2 \right]\,.
\end{align}
The dimensionless coefficient~$\alpha$ will act as a bookkeeping device to do perturbation theory in~$\mathcal{O}(\phi)$. Let us stress that we work at the same order in the tidal response, {\it i.e.}, we neglect non-linear terms arising when one goes beyond linear response theory, as discussed in the previous Section.

The addition of the operator~$\mathcal{O}(\phi)$ in the action modifies the scalar equation of motion, which now schematically reads
\be
\qquad \Box \phi  = r_h^{-2} {\cal S} (\phi) \,,
\ee
where the source term~${\cal S} (\phi)$ is derived from~$\mathcal{O}(\phi)$, and the overall factor of~$r_h^{-2}$ is included for convenience. Notice that we are treating the scalar field as a test perturbation on a fixed black hole background. This implies that we are neglecting the role of the scalar field in the energy momentum tensor when solving for the background Einstein's equations. Our main focus is to investigate the role of interactions in the theory governing the external tidal perturbation on a fixed black hole background.

We will solve the equation of motion in the static limit~$\dot{\phi}=0$ with a decomposition into spherical harmonics,
\be
\phi(\vec{x}) = \sum_{\ell,m} \phi_{\ell m}(r) Y_{\ell m}(\theta,\varphi)\,.
\ee
In the background of a Schwarzschild black hole, the differential equation is of the form
\begin{align} \label{eqn:odesource}
    {\cal L}_\ell \phi_{\ell m}(x) = {\cal S}(\phi_{\ell m}(x)) \,,
\end{align}
where, as before,~$x = r/r_h$, and we have introduced the differential operator
\be
{\cal L}_\ell \equiv \frac{x(x-1)}{x^2} \frac{{\rm d}^2}{{\rm d}x^2} + \frac{2x-1}{x^2} \frac{{\rm d}}{{\rm d}x} - \frac{\ell(\ell+1)}{x^2} \,.
\label{L diff op}
\ee
As the source term~${\cal S} (\phi)$ contains the coupling~$\alpha$, we solve the above equation perturbatively:
\be
\phi_{\ell m}(x) = \phi_{\ell m}^{(0)} + \alpha  \phi_{\ell m}^{(1)} + \alpha^2  \phi_{\ell m}^{(2)} + \ldots
\ee
At any order in perturbation theory, the solution to Eq.~\eqref{eqn:odesource} reads (omitting the angular multipole indices)
\begin{align} 
    \phi (x) = b \phi^+(x)  + c \phi^-(x)   + \int^x {\rm d}y\frac{\phi^+(y) \phi^-(x) -\phi^-(y) \phi^+(x) }{ y(y-1)W[\phi^+(y),\phi^-(y)]} y^2 {\cal S}\big(\phi(y)\big)\,,
\end{align}
where~$b$ and~$c$ are constants, and the integral is indefinite. Notice that~${\cal S}\big(\phi(y)\big)$ is the source term evaluated at the lower order solutions. 
The mode functions~$\phi^{\pm}(x)$ are solutions to the homogeneous equation, and are explicitly given by
\begin{align}
\label{linearsol}
\phi^+(x) &= P_\ell \lp  2x-1 \rp\,; \nonumber \\
\phi^-(x) &= Q_\ell \lp 2x-1 \rp\,.
\end{align}
Here~$P_{\ell}$ is the usual Legendre polynomial at integer~$\ell$, and~$Q_{\ell}(x)$ is the Legendre~$Q$ with the choice of branch cut at~$x\in(-1,1)$. Explicitly,
\begin{align} \label{eqn:phi-}
    Q_{\ell}(x) = \frac{1}{2} P_{\ell}(x) \log \lp \frac{x+1}{x-1} \rp - \sum_{k=1}^{\ell} \frac{1}{k} P_{k}(x)P_{\ell -k}(x)\,.
\end{align}
Furthermore, we have defined the Wronskian
\be
W[\phi^+(x),\phi^-(x)] \equiv  \phi^+(x) \partial_{x} \phi^-(x) - \phi^-(x) \partial_{x} \phi^+(x)  = -\frac{1}{2x(x-1)}\,,
\label{wronskian}
\ee
where in the last step we have used Eq.~\eqref{linearsol}. Thus the general perturbative solution simplifies to 
\begin{align} \label{eqn:gensol}
\boxed{    \phi (x) = b \phi^+(x)  + c \phi^-(x)   - 2 \int^x {\rm d} y\Big(\phi^+(y) \phi^-(x) -\phi^-(y) \phi^+(x) \Big)  \, y^2 {\cal S}\big(\phi(y)\big)\,.}
\end{align}
At the zeroth order in~$\alpha$, the solution is of course just the homogeneous one:
\be
\phi^{(0)} (x) = b^{(0)} \phi^+(x) +c^{(0)}  \phi^-(x)\,.
\label{homogeneous solution}
\ee
Imposing regularity at the horizon selects~$\phi^+$ as the only solution, since $ \phi^- \rightarrow \log(r-r_h)$ as~$r \to r_h$. 
In other words, we set~$c^{(0)} = 0$ to obtain
\be
\phi^{(0)} (x) = b^{(0)} \phi^+(x) = b^{(0)} P_\ell (2x-1) \,.
\label{zeroth order solution}
\ee
Asymptotically, the mode functions behave as~$\phi^+(r) \sim r^\ell$ and~$\phi^-(r) \sim r^{-\ell-1}$ for~$r\rightarrow \infty$. Since
only~$\phi^+$ is allowed, and grows asymptotically, this proves the vanishing of the scalar Love number in the free theory.

At first-order in~$\alpha$, the general solution~\eqref{eqn:gensol} gives
\begin{align}
\phi^{(1)} (x) = c^{(1)} \phi^-(x)  - 2 \int^x {\rm d} y\Big(\phi^+(y) \phi^-(x) -\phi^-(y) \phi^+(x) \Big)  \, y^2 {\cal S}\big(\phi^{(0)}(y)\big)\,.
\label{first order integral}
\end{align}
We have set~$b^{(1)} = 0$ without loss of generality, since a~$\phi^+$ contribution would only renormalize the zeroth-order solution. Correspondingly, to avoid unnecessary superscripts, we will simply make the replacement~$b^{(0)} \rightarrow b$ and write the zeroth-order solution in Eq.~\eqref{homogeneous solution} as~$\phi^{(0)}(x) = b \phi^+(x)$. Both the particular solution and~$\phi^-$ in Eq.~\eqref{first order integral} may contain divergent terms at the horizon. In order for the full solution to be regular at the horizon, the divergent terms must cancel each other. This give a condition on the free coefficient~$c^{(1)}$. The presence of a tail after imposing boundary conditions is then interpreted as a tidal Love number. In the next Section we will compute the Love numbers for different interacting scalar field theories.

\section{Non-linearities in the tidal Love numbers: Examples}
\label{examples}

Following the above discussion on non-linearities in the computation of the tidal Love numbers, we now consider interacting scalar theories on a fixed Schwarzschild black hole background and extract the corresponding tidal response. 

In this Section we consider two classes of scalar self-interactions: $i)$~potential interactions of the power-law form~$\phi^n$; and $ii)$~higher-derivative interactions, specifically~$(\partial\phi)^4$. We will find that a tail is generated both in the ~$\phi^n$ and~$(\partial\phi)^4$ cases, arising respectively at second-order and first-order in perturbation theory. In Sec.~\ref{NLSM sec} we will consider a third class of interacting theories, namely the general non-linear sigma model~$G_{IJ}(\phi) \partial \phi^I \partial \phi^J$. We will find that the non-linear sigma model is tail-less to all orders in perturbation theory, for arbitrary target-space metric~$G_{IJ}(\phi)$.

\subsection{Potential power-law interactions ($\mathcal{O}(\phi) \sim \phi^n$)}

Let us first consider the action of a scalar field with a power-law interaction,
\begin{align}
  S = - \int \rd^4 x \sqrt{-g}\left(\frac{1}{2}(\partial \phi)^2 + \alpha \frac{\phi^n}{n} \right)\,,
\end{align}
with corresponding equation of motion 
\be
\Box \phi = \alpha \phi^{n-1}\,.
\ee
As outlined in Sec.~\ref{nonlin bulk}, we solve this equation perturbatively in~$\alpha$, in the static limit ($\dot{\phi} = 0$) and working with a spherical harmonics decomposition.
The zeroth-order solution, which is regular at the horizon, is given by Eq.~\eqref{zeroth order solution}. The first-order solution is given by Eq.~\eqref{first order integral}, which in this case gives
\be
\phi^{(1)} (x) =  c^{(1)} \phi^-(x)  - 2\, b^{n-1} \int^x {\rm d} y\Big(\phi^+(y) \phi^-(x) -\phi^-(y) \phi^+(x) \Big)  \, y^2 \left(\phi^+(y)\right)^{n-1} \,,
\label{phi 1 gen}
\ee
where we have substituted~$\phi^{(0)} = b \phi^+$ within the source term. Crucially, since~$\phi_+(y)$ is a Legendre polynomial, the source term~$\sim \phi_+^{n-1}(y)$ inside the integral is also a polynomial. In other words, the general solution~\eqref{phi 1 gen} is just a suitable linear combination of terms of the form
\be
\phi^{(1)}_k(x)  = c_{k}^{(1)} \phi^-(x)   -2 \int^x \rd y \Big( \phi^+(y) \phi^-(x) -\phi^-(y) \phi^+(x) \Big) y^{2+k}\,,
\label{phi polynomial source}
\ee
for non-negative integers~$k$. As shown in Appendix~\ref{appendixA}, imposing regularity at the horizon fixes the constant~$c_{k}^{(1)}$ to
\begin{align}
    c_{k}^{(1)} = 2 \frac{(k+3-\ell)_{\ell}}{(k+2)_{\ell +1}}\,,
\end{align}
where~$(q)_\ell = \frac{\Gamma(q+\ell)}{\Gamma(m)}$ is the rising Pochhammer symbol. It is straightforward to deduce that there is no tail associated to this solution for any multipole, that is, the non-linear Love numbers vanish at first-order for power-law interactions. The details are given in Appendix~\ref{appendixA}. 

Importantly, however, a tail does manifest itself at order~$\alpha^2$. The reason is simple. The first-order solution includes log terms. When substituted into the source term at second-order, the source integral generates a dilogarithm,~${\rm Li}_2(1-x)$, which has a tail asymptotically.  We will show this explicitly for the simplest cases of quadratic~($n = 2$) and cubic~($n = 3$) terms. 

\vspace{0.3cm}
\noindent {\bf Quadratic case ($n =2$):} Although a mass term does not technically qualify as an interaction, it nicely illustrates all the relevant physics, and has the technical advantage that the source term is linear. Indeed, resurrecting the multipole indices, Eq.~\eqref{phi 1 gen} gives
\be
\phi^{(1)}_{\ell m} (x) = c^{(1)} \phi^-_{\ell m}(x)  - 2 b\int^x {\rm d} y\Big(\phi^+_{\ell m}(y) \phi^-_{\ell m}(x) -\phi^-_{\ell m}(y) \phi^+_{\ell m}(x) \Big)  \, y^2 \phi^+_{\ell m}(y) \,.
\label{phi 1 n=2}
\ee
As advocated, different~$(\ell,m)$ modes do not mix. 

Let us focus on~$(\ell, m) = (2,0)$ for concreteness. The first order solution~\eqref{phi 1 n=2}, which is regular at the horizon,
is explicitly given by
\be
\phi^{(1)}_{2 0} (x) = \frac{b}{105} \Big(45 x^4+15 x^3-44 x^2+8 \left(6 x^2-6 x+1\right) \log x-40 x+24\Big)\,.
\label{phi 1 n=2 ex}
\ee
Since this only features positive powers of~$x$, there is no tail asymptotically. Therefore the first non-linear tidal Love number is zero. However, let us draw our attention to the presence of a logarithmic term in the solution, which may indicate a mixing between source and response. As shown in Appendix~\ref{appendixA}, this term arises  when the coefficient $c^{(1)}$ is chosen to cancel the non-regular term $\log (x-1)$ in the inhomogeneous source with the one contained in $\phi^- (x)$, leaving a~$\log x$ contribution in the solution.
This logarithm is also responsible for generating a tail at next order in perturbation theory. 

The solution at order~$\alpha^2$ is
\be
\phi^{(2)}_{\ell m} (x) = c^{(2)} \phi^-_{\ell m}(x)  - 2 \int^x {\rm d} y\Big(\phi^+_{\ell m}(y) \phi^-_{\ell m}(x) -\phi^-_{\ell m}(y) \phi^+_{\ell m}(x) \Big)  \, y^2 \phi^{(1)}_{\ell m}(y) \,.
\label{phi 2 n=2}
\ee
Notice that~$\phi^{(1)}$, and its logarithmic contribution, now appears under the integral.
Focusing again on~$(\ell, m) = (2,0)$, we obtain 
\begin{eqnarray}
b^{-1} \phi^{(2)}_{2 0} (x) =  -\frac{64}{11025} \left(6 x^2-6 x+1\right) \text{Li}_2(1-x) +  \frac{1}{84} \left(x^6 + \dots\right) +  \frac{8}{245} \left(x^4+ \dots \right) \log x\,, 
 \label{phi 2,0 n=2}
\end{eqnarray}
where the ellipses denote lower, positive powers of~$x$, and $\text{Li}_2 (1-x)$ is the dilogarithm. 
The latter is regular at the horizon, but generates inverse powers of~$x$ asymptotically:
\be
\phi^{(2)}_{2 0} (x) \supset \frac{64\, b}{11025} \left( - \frac{1}{6x} + \frac{1}{24x^2} + \frac{43 - 60 \log x}{1800 x^3} + {\cal O}\left(\frac{1}{x^4}\right) \right) \,.
\label{n=2 tail}
\ee
The relevant term for~$\ell = 2$ is the~$1/x^3$ contribution. The presence of this term shows that the second-order non-linear Love number may be non-zero. 

However, a note of caution should be stressed at this point. At the non-linear level, a mixing between the source and the response is present in the 
full solution~$\phi^{(2)}$. It is therefore unclear whether this tail should be interpreted as a response or a subleading correction to the source.
The situation is cleaner at the first non-linear level, where we can track the presence of a tail from the particular solution or from the homogeneous~$\phi^-$ solution.
A possible resolution to this problem would require analytic continuation, which, however, is not doable for operators of this form.

\vspace{0.3cm}
\noindent {\bf Cubic case ($n =3$):} In this case the source is quadratic in the lower-order solution. At first order, Eq.~\eqref{phi 1 gen} gives
\be
\phi^{(1)}_{\ell m} (x) = c^{(1)} \phi^-_{\ell m}(x)  - 2 b^2 \int^x {\rm d} y\Big(\phi^+_{\ell m}(y) \phi^-_{\ell m}(x) -\phi^-_{\ell m}(y) \phi^+_{\ell m}(x) \Big)  \, y^2 \mathcal{C}^{\ell m}_{\ell_1 m_1\ell_2 m_2} \phi_{\ell_1 m_1}^+ (y) \phi_{\ell_2 m_2}^+ (y) \,,
\label{phi 1 n=3}
\ee
where the Clebsch-Gordan coefficients~$\mathcal{C}^{\ell m}_{\ell_1 m_1\ell_2 m_2}$ enforce the angular momentum selection rule~$\ell = \ell_1 \otimes \ell_2$,
and a sum over the relevant values of $\ell_1,m_1,\ell_2,m_2$ is understood.

To give an example, let us focus on the contribution from the zeroth-order modes~$(\ell_i, m_i) = (2,0)$. According to the selection rules, they generate a response in the sectors
\be
(\ell_1, m_1) \otimes (\ell_2, m_2) = (2,0) \otimes (2,0) = (0,0), (2,0),~\text{and}~(4,0)\,.
\ee
The~$(2,0) \otimes (2,0)$ contributions to these sectors, imposing regularity at the horizon, are given by 
\begin{align}
\phi^{(1)}_{0 0} (x)  &= \frac{b^2}{210 \sqrt{\pi }} \Big(90 x^6-144 x^5+72 x^4-9 x^3+4 x^2+8 x -21 +8 \log x \Big) \,; \nonumber \\
\phi^{(1)}_{2 0} (x)  &= \frac{b^2}{84 \sqrt{5 \pi }} \Big(60 x^6-90 x^5+45 x^4-11 x^2-10 x+6 +2 \left(6 x^2-6 x+1\right) \log x \Big)\,; \nonumber \\
\phi^{(1)}_{4 0} (x)  &= \frac{b^2}{4620 \sqrt{\pi }}\bigg(3240 x^6-2592 x^5+2117 x^4-10504 x^3+11034 x^2-3572 x+277 \nonumber \\
& ~~~~~~~~~~~~~~~~ + 48 \left(70 x^4-140 x^3+90 x^2-20 x+1\right) \log x \bigg)\,.
\end{align}
These have no tail at spatial infinity, {\it i.e.}, the first non-linear tidal Love number is zero. As in the quadratic case, notice the presence of logarithmic terms,
which indicate a possible mixing between source and response, and are responsible for generating a tail at next order. 

The solution at order~$\alpha^2$ is 
\be
\phi^{(2)}_{\ell m} (x) = c^{(2)} \phi^-_{\ell m}(x)  - 4 b \int^x {\rm d} y\Big(\phi^+_{\ell m}(y) \phi^-_{\ell m}(x) -\phi^-_{\ell m}(y) \phi^+_{\ell m}(x) \Big)  \, y^2 \mathcal{C}^{\ell m}_{\ell_1 m_1\ell_2 m_2} \phi_{\ell_1 m_1}^+ (y) \phi_{\ell_2 m_2}^{(1)} (y) \,.
\label{phi 2 n=3}
\ee
Notice that the source term now features~$\phi^+ (y) \phi^{(1)} (y)$. Focusing once again on the zeroth-order modes~$(\ell_i, m_i) = (2,0)$, the second-order solution receives contribution from all the modes generated at first order, {\it i.e.},~$(\ell_2, m_2) = (0,0), (2,0), (4,0)$. The resulting~$(\ell, m) = (2,0)$ mode function, obtained by summing these several contributions and imposing regularity, is given by 
\begin{align}
b^{-3} \phi^{(2)}_{2 0} (x) =  -\frac{17}{5082 \pi}\left(6 x^2-6 x+1\right) \text{Li}_2(1-x)  + \frac{285}{4004\pi} \left(x^{10} + \dots \right) + \frac{48}{847\pi} \left(x^{8} + \dots \right)\log  x   \,,
\label{phi 2,0 n=3}
\end{align}
where the ellipses denote lower, positive powers of~$x$. The important piece, as in Eq.~\eqref{phi 2,0 n=2}, is $\text{Li}_2(1-x)$, which once again generates a~$1/x^3$ tail near spatial infinity:
\be
\phi^{(2)}_{2 0} (x) \supset \frac{17\, b^3}{5082 \pi} \left( - \frac{1}{6x} + \frac{1}{24x^2} + \frac{43 - 60 \log x}{1800 x^3} + {\cal O}\left(\frac{1}{x^4}\right) \right) \,.
\label{n=3 tail}
\ee
Similarly to the quadratic case, the second-order non-linear Love number may be different from zero, even though this interpretation lacks support due to the possible mixing between source and response.

\subsection{Derivative interactions ($\mathcal{O}(\phi) \sim (\partial \phi)^4$)}

As our second class of examples, consider a theory with higher-derivative interactions,
\begin{align}
  S = - \int \rd^4 x \sqrt{-g}\left[\frac{1}{2}(\partial \phi)^2 + \frac{\alpha}{4} (\partial \phi)^4  \right]\,,
\end{align}
with corresponding equation of motion given by
\be
\Box \phi = - \alpha (\partial \phi)^2 \Box \phi - 2 \alpha \partial^\mu \phi \partial^\nu \phi \nabla_\mu \nabla_\nu \phi \,.
\label{eom dphi4 1}
\ee
As before, we solve this equation perturbatively in~$\alpha$, in the static limit ($\dot{\phi} = 0$) and working with a spherical harmonics decomposition.
In this case we will find that a tail is generated already at first order. 

The zeroth-order solution, regular at the horizon, is given by Eq.~\eqref{zeroth order solution}. Since~$\Box\phi^{(0)} = 0$, Eq.~\eqref{eom dphi4 1} at first order reduces to
\be
\Box \phi^{(1)} = - 2 \alpha \partial^\mu  \phi^{(0)} \partial^\nu \phi^{(0)} \nabla_\mu \nabla_\nu \phi^{(0)} \,.
\label{eom dphi4 2}
\ee
As shown in Appendix~\ref{source term dphi4}, by expanding in spherical harmonics and using the properties of the Schwarzschild background, this equation can be simplified to
\begin{align}
{\cal L}_\ell \phi_{\ell m}^{(1)} &=  - 2 \alpha b^3 \lp f^2 \mathcal{C}^{\ell m}_{\ell_1 m_1\ell_2 m_2 \ell_3 m_3} \,\phi_{\ell_1 m_1}^{+\,'}  \left(\phi_{\ell_3 m_3}^{+\,''} + \frac{f'}{2f}\phi_{\ell_3 m_3}^{+\,'}\right) + 2 \frac{f}{r^2}\, \mathcal{C}^{V_{13} \ell m}_{\ell_1 m_1\ell_2 m_2 \ell_3 m_3}\, \phi_{\ell_1 m_1}^+ \phi_{\ell_3 m_3}^{+\,'} \rp \phi_{\ell_2 m_2}^{+\,'} \nonumber \\
&  ~~~ - 2 \alpha b^3\Bigg(\frac{f}{r^3} \mathcal{C}^{V_{12} \ell m}_{\ell_1 m_1,\ell_2 m_2, \ell_3 m_3}\phi_{\ell_1 m_1}  \phi_{\ell_3 m_3}^{+\,'}  + \frac{1}{r^4}\, \mathcal{C}^{T_{13,23} \ell m}_{\ell_1 m_1\ell_2 m_2 \ell_3 m_3}\, \phi_{\ell_1 m_1}^+  \phi_{\ell_3 m_3}^+ \Bigg) \phi_{\ell_2 m_2}^+ \,,
\label{Box phi 1 dphi4}
\end{align}
where~${\cal L}_\ell$ is defined in Eq.~\eqref{L diff op}, and the scalar, vector and tensor Clebsch-Gordan coefficients are defined in Appendix~\ref{source term dphi4} (see Eqs.~\eqref{CVT}). Thus the source term is characterized by different contributions, which are all responsible for generating a tail in the solution. 

To give an example, consider the contribution from the zeroth-order modes~$(\ell_i, m_i) = (2,0)$. At first order in~$\alpha$ we expect the generation of modes
\be
(\ell_1, m_1) \otimes (\ell_2, m_2)\otimes (\ell_3, m_3) = (2,0)\otimes (2,0)\otimes (2,0) = (0,0), (2,0),(4,0),~\text{and}~(6,0)\,.
\ee
Focusing on the response in the mode~$(\ell,m) = (2,0)$, the solution at order $\alpha$, after imposing proper boundary conditions, reads
\begin{align}
b^{-3} \phi_{2 0}^{(1)} (x) = &  -\frac{540 }{7 \pi }(6 x^2- 6x+1) \big(2 \text{Li}_2(1-x)+\log^2 x\big) - \frac{17820}{49\pi x} \left(x^{5} + \dots \right) \nonumber \\
& -\frac{810}{49 \pi }  \left(30 x^2-86 x+33\right) \log x\,.
\end{align}
The dilogarithm $\text{Li}_2(1-x)$ is once again responsible for generating a~$1/x^3$ tail near spatial infinity:
\be
\phi_{2 0}^{(1)} (x) \supset \frac{b^3}{7 \pi} \left( - \frac{135}{x} + \frac{45}{2x^2} + \frac{3(43 - 60 \log x)}{5 x^3} + {\cal O}\left(\frac{1}{x^4}\right) \right) \,.
\ee
Therefore the first non-linear Love number does not vanish for higher-derivative interacting theories.

Given the examples shown in this Section, we conclude that the number of derivatives present in the relevant operator describing the scalar field interactions seems to provide a good indication of the vanishing or presence of a tail in the field solution. In particular, from the proof briefly described above and shown in Appendix A, a number of derivatives larger than two seems to generate a source term in the equation of motion~$\Box \phi = r_h^{-2} {\cal S} (\phi) \sim x^{k}$, with~$k < 2$, that gives rise to a non-vanishing tail already at first order. For potential interactions, the equation of motion leads to logarithmic terms at first order, which results in a tail at second order.

\section{Non-linear sigma model}
\label{NLSM sec}

As our final example case, we consider non-linear sigma models:
\begin{align} \label{eqn:sigmamodel}
  S = - \int {\rm d}^4 x \sqrt{-g}\left[\frac{1}{2}G_{IJ}(\phi) \partial_{\mu} \phi^I \partial^{\mu} \phi^J  \right]\,.
\end{align}
From a quantum field theory perspective, this class of theories represents the closest scalar analogue to General Relativity.
Indeed, like the Einstein-Hilbert term,~$G_{IJ}(\phi) \partial \phi^I \partial \phi^J$ contains exactly two derivatives and, in principle, all powers of the field. Like General Relativity, the non-linear sigma model admits a geometric interpretation. The connection can be made quite precise,
since Einstein's equations contain a~${\rm O}(2,1)$~$\sigma$ model, irrespective of the symmetries of the background~\cite{Sanchez:1981mn}.
As we will show in this Section, this class of interacting theory has exactly vanishing Love number to all orders in perturbation theory.

For the example to be non-trivial, it is of course essential that the metric function~$G_{IJ}$ in general describes a curved target space, such that the theory is not reducible to decoupled free fields. By choosing a Riemann normal coordinate at~$\phi^I=0$, the target space metric can be expanded as
\begin{align} \label{eqn:eommulti}
    G_{IJ} = \delta_{IJ} -\frac{1}{3}R_{I KJL}(0)\phi^{K}\phi^{L} + {\cal O}(\phi^3)\,.
\end{align}
We will focus on the perturbative regime in which the Riemann tensor~$R_{I KJL}(0)$ and its covariant derivatives~$\nabla \cdots \nabla R_{I KJL}(0)$ are small and can be treated perturbatively. As long as the field space is not Riemann flat, this is an interacting theory.

The equation of motion for~$\phi^I$ is simply
\begin{align} \label{eqn:sigmamodeleqn}
    \Box \phi^{I} + \Gamma^I_{KL}(\phi) \partial_{\mu} \phi^K \partial^{\mu} \phi^L =0\,,
\end{align}
where~$ \Gamma^I_{KL}$ denotes the Christoffel symbols associated with~$G_{IJ}$. Therefore, by expanding~$\Gamma^I_{KL}$ around the origin using the Riemann normal coordinate, the perturbative series solution can be defined.  

Remarkably, we will prove that this model has {\it no tail at all orders in perturbation theory}. The detailed proof, given in Sec.~\ref{app:proofsigmamodel},
can be understood intuitively as follows. If the field space is Riemann flat, then the theory can be reduced to a number of decoupled free scalars, which we already know are tail-less. On the other hand, one can choose field-space coordinates such that the flat metric is expressed in some non-trivial function of~$\phi^I$. Since this is just a field redefinition of the free scalars, the solution in this basis also has no tail, as long as the field redefinition can be done perturbatively. Now, consider the more general case of a non-flat target space. The key observation is that, when Eq.~\eqref{eqn:sigmamodeleqn} is solved perturbatively, there is no distinction between a curved field space and a flat field space with a non-trivial metric. Therefore there is no tail in general.

\subsection{Explicit example} 

Before giving the general proof, it is instructive to work out an explicit example with two scalar fields~$\phi$ and~$\chi$.
Consider the interacting theory
\begin{align}
  S = - \int \rd^4 x \sqrt{-g}\left[\frac{1}{2}\big(1 + 2\beta \phi\big)(\partial \phi)^2 + \frac{1}{2}(\partial \chi)^2 + \alpha \phi \chi \partial_\mu \phi  \partial^\mu \chi  \right]\,,
\end{align}
corresponding to the field space metric
\be
G_{\phi\phi} = 1 + 2\beta \phi\,;\qquad G_{\phi\chi} = G_{\chi\phi} = \alpha \phi \chi  \,;\qquad G_{\chi\chi} = 1\,.
\ee
It is easy to check that this metric is not Riemann flat, as long as~$\alpha\neq 0$. The coefficient~$\beta$ explicitly breaks the exchange symmetry~$\phi \longleftrightarrow \chi$, and ensures that the absence of tails demonstrated below is not an artefact of this symmetry. 

The equations of motion can be expressed as
\begin{align}
\Box \phi & =\frac{\alpha ^2 \chi ^2 \phi -\beta }{1-\alpha ^2 \chi ^2 \phi ^2+2 \beta  \phi } (\partial \phi)^2 + \frac{\alpha  \phi }{1-\alpha ^2 \chi ^2 \phi ^2+2 \beta  \phi } (\partial \chi)^2  \nonumber \\
\Box \chi & = \frac{\alpha \beta \chi   \phi +\alpha \chi }{1-\alpha ^2 \chi ^2 \phi ^2+2 \beta  \phi } (\partial \phi)^2 - \frac{\alpha ^2 \chi  \phi ^2}{1-\alpha ^2 \chi ^2 \phi ^2+2 \beta  \phi }(\partial \chi)^2 \,.
 \end{align}
As before, we work in the static limit~($\dot{\phi} = \dot{\chi} = 0$) and decompose the fields in spherical harmonics. Furthermore,
we solve the equations perturbatively in~$\alpha$ and~$\beta$, treating these as the same order in the expansion.
In other words, the fields are expanded perturbatively as
\begin{align}
\phi & = \phi^{(0)} +  \phi^{(1)} + \phi^{(2)}  + \dots\,; \nonumber \\
\chi & = \chi^{(0)} + \chi^{(1)} + \chi^{(2)} + \dots\,,
\end{align}
where $\phi^{(1)} = \alpha \phi^{(1,0)} +\beta \phi^{(0,1)}$, $\phi^{(2)} = \alpha^2 \phi^{(2,0)} + \beta^2 \phi^{(0,2)}+ \alpha \beta \phi^{(1,1)}$, etc.  

The zeroth-order equations are just~$\Box \phi^{(0)} = \Box \chi^{(0)} = 0$, with regular solution given as before by
\be
\phi^{(0)}(x) = \chi^{(0)}(x) = b \phi^+(x)\,.
\ee
At first order in~$\alpha$ and $\beta$, the equations of motion simplify to
\begin{align}
\Box \phi^{(1)} & = - \alpha \phi^+ \big(\partial \chi^{(0)}\big)^2 - \beta \big(\partial \phi^{(0)}\big)^2\,; \nonumber \\
\Box \chi^{(1)} & = - \alpha \chi^{(0)} \big(\partial \phi^{(0)}\big)^2\,.
\label{EOM sigma 1st}
\end{align}
As shown in Appendix~\ref{source term sigma model}, expanding in spherical harmonics and using the properties of the Schwarzschild background, these equations can be simplified to
\begin{align}
{\cal L}_\ell \phi_{\ell m}^{(1)} = &  - \alpha b^3 \lp  f \phi_{\ell_1 m_1}^{+ \, '} \phi_{\ell_2 m_2}^{+ \, '}  \mathcal{C}^{\ell m}_{\ell_1 m_1,\ell_2 m_2, \ell_3 m_3} + \frac{1}{r^2} \phi_{\ell_1 m_1}^{+} \phi_{\ell_2 m_2}^{+} \mathcal{C}^{V_{12} \ell m}_{\ell_1 m_1,\ell_2 m_2, \ell_3 m_3}\rp \phi_{\ell_3 m_3}^{+} \nonumber \\
& - \beta b^2 \lp  f \phi_{\ell_1 m_1}^{+ \, '} \phi_{\ell_2 m_2}^{+ \, '} \mathcal{C}^{\ell m}_{\ell_1 m_1,\ell_2 m_2} + \frac{1}{r^2} \phi_{\ell_1 m_1}^{+} \phi_{\ell_2 m_2}^{+} \mathcal{C}^{V_{12} \ell m}_{\ell_1 m_1,\ell_2 m_2} \rp \,;
\nonumber \\
{\cal L}_\ell \chi_{\ell m}^{(1)}  = & - \alpha b^3  \lp  f \phi_{\ell_1 m_1}^{+ \, '} \phi_{\ell_2 m_2}^{+ \, '}  \mathcal{C}^{\ell m}_{\ell_1 m_1,\ell_2 m_2, \ell_3 m_3} + \frac{1}{r^2} \phi_{\ell_1 m_1}^{+} \phi_{\ell_2 m_2}^{+}  \mathcal{C}^{V_{12} \ell m}_{\ell_1 m_1,\ell_2 m_2, \ell_3 m_3} \rp \phi_{\ell_3 m_3}^{+}\,,
\label{NL sig example 1st}
\end{align}
in terms of properly defined scalar, vector and tensor Clebsch-Gordan coefficients.
Notice that, since~$\phi$ and~$\chi$ are identical at zeroth-order, the solutions will be symmetric at first-order in~$\alpha$.  However, the presence of the interaction term with coupling $\beta$ breaks this symmetry, leading to different solutions for the two fields. 

Similarly to what was done in the previous examples, let us consider a zeroth-order mode~$(\ell_i, m_i) = (2,0)$, which will generate at first order the following modes due to selection rules~$(\ell, m) = (\ell_1, m_1) \otimes (\ell_2, m_2) \otimes (\ell_3, m_3) = (0,0), (2,0), (4,0), (6,0)$. The corresponding solutions for~$\phi^{(1)}$ are
\begin{align}
\phi^{(1)}_{0 0} (x) = & -\alpha \, b^3 \frac{3 \sqrt{5}}{14 \pi }  (x-1) x \big(12 x^4-24 x^3+18 x^2-6 x+1\big) \nonumber \\
& - \beta \, b^2 \frac{3}{\sqrt{\pi }}  (x-1) x \left(3 x^2-3 x+1\right) \,;
\nonumber \\
\phi^{(1)}_{2 0} (x)  = & -\alpha \, b^3 \frac{15}{14 \pi }  x \left(3 x^2-3 x+1\right) \left(6 x^3-12 x^2+7 x-1\right) \nonumber \\
& - \beta \, b^2 \frac{3}{7} \sqrt{\frac{5}{\pi }} x \left(6 x^3-12 x^2+7 x-1\right)\,; \nonumber \\
\phi^{(1)}_{4 0} (x)  = & \,\, \alpha \, b^3 \frac{3 \sqrt{5}}{77 \pi} (x-1) x \left(-108 x^4+216 x^3-127 x^2+19 x+1\right) \nonumber \\
& + \beta \, b^2 \frac{3}{7 \sqrt{\pi }} (x-1) x \left(17 x^2-17 x+4\right)\,; \nonumber \\
\phi^{(1)}_{6 0} (x)  = & \,\, \alpha \, b^3 \frac{45}{154 \pi } \sqrt{\frac{5}{13}}  (x-1) x \left(118 x^4-236 x^3+163 x^2-45 x+4\right)\,.
\end{align}
The solutions for $\chi^{(1)}$ are simply obtained by setting~$\beta = 0$ in the above, {\it i.e.}, $\chi^{(1)}_{\ell m} = \phi^{(1)}_{\ell m}\big|_{\beta = 0}$.
Evidently, none of these solutions have a tail at spatial infinity, hence the first-order non-linear Love number vanishes for this multi-field target space theory.
Furthermore, in contrast with the power-law examples (see Eqs.~\eqref{phi 1 n=2} and~\eqref{phi 1 n=3}), notice that the first-order solutions do not have any~$\log x$ term.
This suggests that dilogarithms will not be generated at second order, and we will confirm that this is indeed the case.

At the next-to-leading order, that is at order~$\alpha^2$, $\beta^2$ and $\alpha \beta$, the equations of motion become
\begin{align}
\Box \phi^{(2)} & =  \alpha^2 \phi^{(0)} \chi^{(0) \,2} \big(\partial \phi^{(0)}\big)^2 - \alpha \phi^{(1)} \big(\partial \chi^{(0)}\big)^2 
- 2\alpha \phi^{(0)} \partial^\mu \chi^{(0)} \partial_\mu \chi^{(1)} \nonumber \\
& ~~~ + 2 \alpha \beta \phi^{(0)\, 2} \big(\partial \chi^{(0)}\big)^2 + 2 \beta^2 \phi^{(0)} \big(\partial \phi^{(0)}\big)^2 - 2 \beta \partial^\mu \phi^{(0)}\partial_\mu \phi^{(1)}\,, \nonumber \\
\Box \chi^{(2)} & =  \alpha^2 \chi^{(0)} \phi^{(0)\, 2} \big(\partial \chi^{(0)}\big)^2  - \alpha \chi^{(1)} \big(\partial \phi^{(0)}\big)^2 
- 2\alpha \chi^{(0)} \partial^\mu \phi^{(0)} \partial_\mu \phi^{(1)} \nonumber \\
& ~~~ + \alpha \beta \phi^{(0)} \chi^{(0)} \big(\partial \phi^{(0)}\big)^2\,.
\label{EOM sigma 2nd}
\end{align}
As shown in Appendix~\ref{source term sigma model}, these can be rewritten as
\begin{align}
{\cal L}_\ell \phi^{(2)}_{\ell m} & =  \alpha^2 b^5  \phi^+_{\ell_3 m_3} \phi^+_{\ell_4 m_4} \phi^+_{\ell_5 m_5}\lp f  \phi_{\ell_1 m_1}^{+\,'} \phi_{\ell_2 m_2}^{+\, '} \mathcal{C}^{\ell m}_{\ell_1 m_1,\ldots,\ell_5 m_5} +  \frac{1}{r^2}  \phi_{\ell_1 m_1}^{+} \phi_{\ell_2 m_2}^{+} \mathcal{C}^{V_{12} \ell m}_{\ell_1 m_1,\ldots, \ell_5 m_5} \rp \nonumber \\
& - \alpha b^2 \phi^{(1)}_{\ell_3 m_3} \lp f \phi_{\ell_1 m_1}^{+\, '} \phi_{\ell_2 m_2}^{+\, '} \mathcal{C}^{\ell m}_{\ell_1 m_1,\ell_2 m_2, \ell_3 m_3}
+ \frac{1}{r^2}  \phi_{\ell_1 m_1}^{+} \phi_{\ell_2 m_2}^{+}  \mathcal{C}^{V_{12} \ell m}_{\ell_1 m_1,\ell_2 m_2, \ell_3 m_3} \rp
\nonumber \\
& - 2\alpha b^2  \phi^{+}_{\ell_3 m_3} \lp f  \chi_{\ell_1 m_1}^{(1)\, '} \phi_{\ell_2 m_2}^{+\, '} \mathcal{C}^{\ell m}_{\ell_1 m_1,\ell_2 m_2, \ell_3 m_3}
+ \frac{1}{r^2} \chi_{\ell_1 m_1}^{(1)} \phi_{\ell_2 m_2}^+  \mathcal{C}^{V_{12} \ell m}_{\ell_1 m_1,\ell_2 m_2, \ell_3 m_3} \rp
\nonumber \\
& + 2 \alpha \beta b^4 \phi^{+}_{\ell_3 m_3} \phi^{+}_{\ell_4 m_4} \lp f  \phi_{\ell_1 m_1}^{+\, '} \phi_{\ell_2 m_2}^{+\, '}  \mathcal{C}^{\ell m}_{\ell_1 m_1,\ldots, \ell_4 m_4}
+ \frac{1}{r^2}  \phi_{\ell_1 m_1}^{+} \phi_{\ell_2 m_2}^{+}  \mathcal{C}^{V_{12},\ell m}_{\ell_1 m_1,\ldots, \ell_4 m_4} \rp
\nonumber \\
& + 2 \beta^2 b^3 \phi^{+}_{\ell_3 m_3} \lp f \phi_{\ell_1 m_1}^{+\,'} \phi_{\ell_2 m_2}^{+\, '}  \mathcal{C}^{\ell m}_{\ell_1 m_1,\ell_2 m_2, \ell_3 m_3}
+ \frac{1}{r^2} \phi_{\ell_1 m_1}^{+} \phi_{\ell_2 m_2}^{+}  \mathcal{C}^{V_{12},\ell m}_{\ell_1 m_1,\ell_2 m_2, \ell_3 m_3} \rp
\nonumber \\
& - 2\beta b \lp f  \phi_{\ell_1 m_1}^{(1)\, '} \phi_{\ell_2 m_2}^{+\, '}  \mathcal{C}^{\ell m}_{\ell_1 m_1,\ell_2 m_2}
+ \frac{1}{r^2} \phi_{\ell_1 m_1}^{(1)} \phi_{\ell_2 m_2}^{+}   \mathcal{C}^{V_{12},\ell m}_{\ell_1 m_1,\ell_2 m_2} \rp\,,
\label{NL sig example 2nd a}
\end{align}
and
\begin{align}
{\cal L}_\ell \chi^{(2)}_{\ell m} & = \alpha^2 b^5  \phi^+_{\ell_3 m_3} \phi^+_{\ell_4 m_4} \phi^+_{\ell_5 m_5}\lp f  \phi_{\ell_1 m_1}^{+\,'} \phi_{\ell_2 m_2}^{+\, '} \mathcal{C}^{\ell m}_{\ell_1 m_1,\ldots,\ell_5 m_5} +  \frac{1}{r^2}  \phi_{\ell_1 m_1}^{+} \phi_{\ell_2 m_2}^{+} \mathcal{C}^{V_{12} \ell m}_{\ell_1 m_1,\ldots, \ell_5 m_5} \rp \nonumber \\
& - \alpha b^2 \chi^{(1)}_{\ell_3 m_3} \lp f \phi_{\ell_1 m_1}^{+ \, '} \phi_{\ell_2 m_2}^{+\, '} \mathcal{C}^{\ell m}_{\ell_1 m_1,\ell_2 m_2, \ell_3 m_3}
+ \frac{1}{r^2}  \phi_{\ell_1 m_1}^{+} \phi_{\ell_2 m_2}^{+}  \mathcal{C}^{V_{12} \ell m}_{\ell_1 m_1,\ell_2 m_2, \ell_3 m_3} \rp
\nonumber \\
& - 2\alpha b^2 \phi^{+}_{\ell_3 m_3} \lp f  \phi_{\ell_1 m_1}^{(1)\, '} \phi_{\ell_2 m_2}^{+\, '} \mathcal{C}^{\ell m}_{\ell_1 m_1,\ell_2 m_2, \ell_3 m_3}
+ \frac{1}{r^2} \phi_{\ell_1 m_1}^{(1)} \phi_{\ell_2 m_2}^{+}  \mathcal{C}^{V_{12} \ell m}_{\ell_1 m_1,\ell_2 m_2, \ell_3 m_3} \rp
\nonumber \\
& + \alpha \beta b^4 \phi^{+}_{\ell_3 m_3} \phi^{+}_{\ell_4 m_4} \lp f  \phi_{\ell_1 m_1}^{+\, '} \phi_{\ell_2 m_2}^{+\, '}  \mathcal{C}^{\ell m}_{\ell_1 m_1,\ldots, \ell_4 m_4} + \frac{1}{r^2}  \phi_{\ell_1 m_1}^{+} \phi_{\ell_2 m_2}^{+}  \mathcal{C}^{V_{12},\ell m}_{\ell_1 m_1,\ldots, \ell_4 m_4} \rp\,.
\label{NL sig example 2nd b}
\end{align}
Following the example above, we can investigate the response in the mode~$(\ell,m) = (2,0)$.
Taking into account the contributions coming from the various mixings due to the angular momentum selection rules, we get
\begin{align}
\phi^{(2)}_{20} (x) & = (x-1) x\left(6 x^2-6 x+1\right)\Bigg\{ \beta^2 b^3 \frac{60}{7 \pi }  \left(3 x^2-3 x+1\right)  \nonumber \\
& + \alpha \beta b^4 \frac{15 \sqrt{5}}{154 \pi ^{3/2}}  \Big(279 x^4-558 x^3+411 x^2-132 x+20\Big) \nonumber \\
& +  \alpha^2 b^5 \frac{75}{8008 \pi ^2}   \Big(13167 x^6-39501 x^5+48171 x^4-30507 x^3  +10743 x^2-2073 x+212\Big)\Bigg\}\,; \nonumber \\
\chi^{(2)}_{20} (x) & =  (x-1) x\left(6 x^2-6 x+1\right)\Bigg\{ \alpha \beta b^4 \frac{15 \sqrt{5}}{154 \pi ^{3/2}}  \left(132 x^4-264 x^3+195 x^2-63 x+10\right) \nonumber \\
& + \alpha^2 b^5 \frac{75}{8008 \pi ^2}  \Big(13167 x^6-39501 x^5+48171 x^4-30507 x^3 +10743 x^2-2073 x+212\Big) \Bigg\}\,.
\end{align}
These have no tail, hence the second-order non-linear Love number vanishes. Furthermore, notice once again the absence of any~$\log x$ term, which
suggests that dilogarithms will not be generated at the next order. 

\subsection{Proof: non-linear sigma models generate no tail to all orders in perturbation theory} 
\label{app:proofsigmamodel}

Let us now prove that the non-linear sigma model~\eqref{eqn:sigmamodel} generates no tail at all orders in perturbation theory, for arbitrary target-space metric. 

The starting point is to consider a single field model, with target space metric~$G(\phi)$. The equation of motion reads
\begin{align}
    \Box \phi  = - \frac{1}{2}G(\phi)^{-1}\partial_{\phi}G(\phi) (\partial\phi)^2\,. 
\end{align}
Provided that the field space metric is non-singular, one can always find a field redefinition~$\phi(\chi)$, such that~$\chi$ solves~$\Box \chi =0$. Since the solution for~$\chi$ has no tail at any order, the perturbative series of~$\phi$, assuming that~$G(\phi)$ is sufficiently regular for the series to exist, can be obtained readily and will have no tail. 

Let us take a simple example with one single expansion parameter~$\alpha$, and rewrite the equation of motion as 
\begin{align}
     \Box \phi  = \sum_{k=0}^\infty c_k \alpha^{k+1} \phi^{k} (\partial\phi)^2\,.
\end{align}
The perturbative series 
\begin{align}
    \phi = \sum_{i=0}^\infty \alpha^i \phi^{(i)} = \phi^{(0)} + \alpha \phi^{(1)}+ \alpha^2 \phi^{(2)} + \dots 
\end{align}
solves the equation of motion order by order in~$\alpha$ by 
\begin{align} \label{eqn:pertsource}
    \Box \phi^{(n)}  =\sum_{k}  c_k \sum_{\sum_{j}m_j + p+q+k+1 =n}  \,\prod_{j=1}^k\phi^{(m_j)} \partial_{\mu} \phi^{(p)} \partial^{\mu} \phi^{(q)}.
\end{align} 
Since the coefficients~$c_k$ are independent, each source term labeled by~$k$ on the right-hand side, 
\begin{align}
 \sum_{\sum_{j}m_j + p+q+k+1 =n}  \prod_{j=1}^k\phi^{(m_j)} \partial_{\mu} \phi^{(p)} \partial^{\mu} \phi^{(q)}\,,
\end{align}
does not give rise to any tail in~$\phi^{(n)}$ individually.
One can also observe that the solution for lower order~$\phi^{(m_i)}$ will depend on~$c_{m_i-1}$, we can then further conclude that each term of the form 
\begin{align}
     \sum_{\sigma}  \prod_{j=1}^k\phi^{(\sigma(j))} \partial_{\mu} \phi^{(\sigma(j+1))} \partial^{\mu} \phi^{(\sigma(j+2))}\,,
\end{align}
with~$\sigma$ being the permutation set of fixed~$\{m_1, \dots, m_k, p, q\}$, does not give rise to any tail in~$\phi^{(n)}$. Furthermore, the~$\phi^{(n)}$ obtained from individual lower-level source terms will not generate tails for~$\left.\phi^{(n')}\right|_{n' >n}$ at higher orders subsequently. 

With the above knowledge at hand, let us return to the multi-field case in Eq.~\eqref{eqn:sigmamodeleqn}. It can be written as 
\begin{align}
    \Box \phi^{I}  = \sum_{i} c^I_{KL|M_1 \dots M_i}\phi^{M_1} \cdots \phi^{M_i} \partial_{\mu} \phi^K \partial^{\mu} \phi^L\,,
\end{align}
as long as the metric~$G_{IJ}(\phi)$ is regular for the series to exist. When this equation is solved perturbatively at~$n^{\rm th}$ order, every term on the right-hand side,
\begin{align}
    \phi^{M_1} \cdots \phi^{M_i} \partial_{\mu} \phi^K \partial^{\mu} \phi^L,
\end{align}
takes the explicit solutions of~$\phi^J$ from lower order and therefore has exactly the same form as Eq.~\eqref{eqn:pertsource}, with explicit solutions of the field substituted in. From the argument of a single field, we can conclude that~$\phi^{I (n)}$ do not have tails at all orders in perturbation theory.

\section{Conclusions}
\label{conclu sec}

Tidal Love numbers describe the tidal response of compact objects under the presence of external perturbations and thus provide insightful information on their structure and interior.
Within Einstein gravity, the Love numbers of black holes are found to be exactly zero for scalar, vector and tensor perturbations.
This result has been obtained assuming linear order in perturbation theory. However, gravity is by itself a non-linear theory, so non-linearities could be important in the estimate of the tidal Love numbers.

In this work we investigated the role of non-linearities in the computation of the tidal Love numbers, focusing on external scalar perturbations on the background of a Schwarzschild black hole. We first studied which kind of non-linearities may be involved in the theory. The first kind is based on going beyond linear response theory, usually assumed when one considers Love numbers. The second, which is the main focus of this work, is in the sector of external perturbations. 

In particular, we studied interacting scalar fields on a black hole background and computed the Love numbers for different types of scalar interactions, treating the interactions in a perturbative expansion. For power-law interactions, we found that the Love numbers are still zero at first order in perturbation theory, but may be different from zero at next-to-next-to-leading order, even though a proper claim should be made only once the issue of source-response mixing is solved. For higher-derivative interactions, we found that the Love numbers are non-vanishing already at first order. Our most interesting class of interacting scalar theories is the non-linear sigma model. Here we found, remarkably, that the Love numbers vanish to all orders in perturbation theory. It would be very interesting to see whether this is the consequence of a symmetry. 

Our work is intended only as a first step in understanding the role of non-linearities in the tidal deformability of black holes and, as such, it can be extended in several ways. Firstly, understanding how to disentangle the source and response at non-linear level is crucial to properly extract Love numbers from the full relativistic solution.
Secondly, it would be interesting to determine the size of the response when going beyond the assumption of linear response theory (the first kind of non-linearity we mention) compared to the ones arising from the theory. Finally, while we focused on the interesting case of interacting scalar perturbations around black holes, in principle external tensor perturbations are the relevant one for a binary black hole system, which could be detected at present and future gravitational wave experiments.
In this case, Einstein gravity already provides non-linearities in the game, and these have recently received attention in the context of black hole quasi-normal modes and ringdown~\cite{Mitman:2022qdl,Cheung:2022rbm,Ma:2022wpv,Lagos:2022otp,Kehagias:2023ctr,Baibhav:2023clw,Kehagias:2023mcl}.
We plan to investigate these issues in future work.

\subsubsection*{Acknowledgments}
We thank M.~Ivanov, A.~Joyce, U.~Kol and A.~Riotto for interesting comments and discussions.
V.DL. is supported by funds provided by the Center for Particle Cosmology at the University of Pennsylvania. 
The work of J.K. and S.W. is supported in part by the DOE (HEP) Award DE-SC0013528.

\begin{appendices}

\section{Vanishing tail for source of the form~$y^k$,~$k\ge-2$, in Eq.~\eqref{phi polynomial source}} 
\label{appendixA}

We give some analytic properties for the solution \eqref{phi polynomial source} with the source of the form~${\cal S}\sim y^k$ for integers~$k$, 
\begin{align}
\phi^{(1)}_k(x)  = c_{k}^{(1)} \phi^-(x)   -2 \int^x \rd y \Big( \phi^+(y) \phi^-(x) -\phi^-(y) \phi^+(x) \Big) y^{2+k}\,,
\end{align}
From the explicit expression of~$\phi^-$ in Eqs.~\eqref{linearsol} and~\eqref{eqn:phi-}, 
\begin{align}
    \phi^-(x) = \frac{1}{2} \phi^+(x) \log \left( \frac{x}{x-1}\right) + \sum_j b_j x^j\,,
\label{phi- append}
\end{align}
we see that only the first term~$\frac{1}{2} \phi^+(x) \log \left( \frac{x}{x-1}\right)$ is relevant for any fall-off tail at large~$r$. Therefore, in the inhomogeneous part we can focus on 
\begin{align}
&- \int^x \rd y \Bigg[ \phi^+(y) \phi^+(x) \log \left( \frac{x}{x-1}\right) -\phi^+(y) \log \left( \frac{y}{y-1}\right) \phi^+(x) \Bigg] y^{2+k} \nonumber \\
&\qquad =  \phi^+(x) \Bigg[  - \log \left( \frac{x}{x-1}\right)\int^x \rd y \,\phi^+(y) y^{2+n} + \int^x {\rm d}y \,\phi^+(y) \log \left( \frac{y}{y-1}\right)  y^{2+k} \Bigg] \nonumber \\ 
& \qquad = -\phi^+(x) \int^x \rd y \left(\int^y \rd z\, \phi^+(z) z^{2+k}  \right) \frac{{\rm d}}{{\rm d}y} \log \left( \frac{y}{y-1} \right)\nonumber \\ 
& \qquad =  \phi^+(x)\int^x \rd y \Bigg(  \sum_{m=2+k,\,m\neq -1}^{2+k+\ell} \frac{a_m y^{m+1}}{m+1}  + a_{-1} \log y \Bigg) \frac{1}{y(y-1)} \nonumber \\
& \qquad =  \phi^+(x)\Bigg[\sum_{m=2+k}^{-2}\frac{a_m}{m+1} \Bigg(\log \left( \frac{x-1}{x}\right)-\sum_{j=m}^{-1}\frac{x^j}{j}   \Bigg) -a_{-1}\left( \text{Li}_2(1-x)+\frac{1}{2} \log^2 x\right)  \nonumber \\
& \qquad ~~~~~~~~~~~~~ +  \sum_{m={\rm max}\{0, 2+k \}}^{2+n+\ell} \frac{a_m}{m+1} \Bigg( \log(x-1)+\sum_{j=1}^m\frac{x^j}{j} \Bigg) \Bigg]\,,
\end{align}
 where in the third equality we have used the fact that~$\phi^+(z) z^{2+k}$ can always be written in the form~$\sum_{m} a_m z^{m}$. We therefore find that only when~$k\ge-2$ there is no powers of~$\frac{1}{x}$, and the coefficient~$c^{(1)}_k$ is also fixed by the regular boundary condition at the horizon,
 \begin{align}
     c^{(1)}_k = 2  \sum_{m = 2+k}^{2+k+\ell} \frac{a_m}{m+1}  = 2 \frac{(k+3-\ell)_{\ell}}{(k+2)_{\ell+1}}\,,
\label{c1 append}
 \end{align}
 where~$(q)_{\ell}= \frac{\Gamma(q+\ell)}{\Gamma(q)}$ is the rising factorial.

\section{Simplification of the source term for~$(\partial\phi)^4$}
\label{source term dphi4}

In this Appendix we show how the source term for the~$(\partial\phi)^4$ interaction can be simplified.
Equation~\eqref{eom dphi4 2} for the first order field is
\be
\Box \phi^{(1)} = - 2 \alpha \partial^\mu  \phi^{(0)} \partial^\nu \phi^{(0)} \nabla_\mu \nabla_\nu \phi^{(0)} \,.
\ee
By expanding the field in spherical harmonics, the derivative and two gradients terms become
\begin{align}
\partial^a \phi &= g^{a c} \partial_c \phi = g^{a c} \big(Y_{\ell m} \partial_c \phi_{\ell m} + \phi_{\ell m} Y_{\ell m;c}\big)\,; \nonumber \\
\nabla_b \nabla_a \phi &= Y_{\ell m} \nabla_b \nabla_a \phi_{\ell m}  + Y_{\ell m;b} \partial_a  \phi_{\ell m} + Y_{\ell m, a} \partial_b \phi_{\ell m} + \phi_{\ell m} Y_{\ell m; b a}\,,
\end{align}
where we have used the compact notation~$\nabla_a Y_{\ell m} = \partial_a Y_{\ell m} \equiv Y_{\ell m; a}$. Putting everything together, we get
\begin{align}
\left({\cal L}_\ell \phi_{\ell m}^{(1)}\right) Y_{\ell m} = & - 2 \alpha b^3 g^{a c} g^{b d} \Big(Y_{\ell_1 m_1} Y_{\ell_2 m_2} \partial_c \phi_{\ell_1 m_1}^+  \partial_d \phi_{\ell_2 m_2}^+ +  Y_{\ell_2 m_2} Y_{\ell_1 m_1;c} \phi_{\ell_1 m_1}^+ \partial_d \phi_{\ell_2 m_2}^+ \nonumber \\
& ~~~~~~~~~~~~~~~~~~ +  Y_{\ell_1 m_1} Y_{\ell_2 m_2;d} \partial_c \phi_{\ell_1 m_1}^+ \phi_{\ell_2 m_2}^+  + Y_{\ell_1 m_1;c} Y_{\ell_2 m_2;d} \phi_{\ell_1 m_1}^+ \phi_{\ell_2 m_2}^+\Big) \nonumber \\
& \times  \Big(Y_{\ell_3 m_3} \nabla_b \nabla_a \phi_{\ell_3 m_3}^+  + Y_{\ell_3 m_3;b} \partial_a  \phi_{\ell_3 m_3}^+ + Y_{\ell_3 m_3; a} \partial_b \phi_{\ell_3 m_3}^+ +  Y_{\ell_3 m_3; b a}\phi_{\ell_3 m_3}^+\Big)\,.
\end{align}
For a Schwarzschild background, some of these terms are null. What is left is
\begin{align}
\left({\cal L}_\ell \phi_{\ell m}^{(1)}\right) Y_{\ell m} = &- 2 \alpha b^3  
\Big( Y_{\ell_1 m_1} Y_{\ell_2 m_2} Y_{\ell_3 m_3} g^{rr} g^{rr} \phi_{\ell_1 m_1}^{+\,'}  \phi_{\ell_2 m_2}^{\,'} \nabla_r \nabla_r \phi_{\ell_3 m_3}^+  \nonumber \\
& ~~~~~~~~~~ + 2 g^{ac} Y_{\ell_2 m_2} Y_{\ell_3 m_3; a} Y_{\ell_1 m_1;c} g^{rr} \phi_{\ell_1 m_1}^+  \phi_{\ell_2 m_2}^{+\,'} \phi_{\ell_3 m_3}^{+\,'} \nonumber \\
& ~~~~~~~~~~ + g^{a c} g^{b d} Y_{\ell_1 m_1;c} Y_{\ell_2 m_2;d} Y_{\ell_3 m_3} \phi_{\ell_1 m_1}^+ \phi_{\ell_2 m_2}^+ \nabla_b \nabla_a \phi_{\ell_3 m_3}^+  \nonumber \\
& ~~~~~~~~~~  + g^{a c} g^{b d} Y_{\ell_1 m_1;c} Y_{\ell_2 m_2;d} Y_{\ell_3 m_3; b a} \phi_{\ell_1 m_1}^+ \phi_{\ell_2 m_2}^+ \phi_{\ell_3 m_3}^+ \Big)\,,
\label{Boxphi 1 append B}
\end{align}
Using~$g^{rr} = f$, together with
\begin{align}
\nabla_r \nabla_r \phi_{\ell_3 m_3}^+ &= \phi_{\ell_3 m_3}^{+\,''} - \Gamma^r_{rr} \phi_{\ell_3 m_3}^{+\,'} = \phi_{\ell_3 m_3}^{+\,''} + \frac{f'}{2f}\phi_{\ell_3 m_3}^{+\,'}\,; \nonumber \\
\nabla_\theta \nabla_\theta \phi_{\ell_3 m_3}^+ &= - \Gamma^r_{\theta \theta} \phi_{\ell_3 m_3}^{+\,'} = r f \phi_{\ell_3 m_3}^{+\,'}\,; \nonumber \\
\nabla_\varphi \nabla_\varphi \phi_{\ell_3 m_3}^+ &= - \Gamma^r_{\varphi \varphi} \phi_{\ell_3 m_3}^{+\,'} = r f \sin^2\theta \phi_{\ell_3 m_3}^{+\,'}\,,
\end{align}
Eq.~\eqref{Boxphi 1 append B} simplifies to
\begin{align}
\left({\cal L}_\ell \phi_{\ell m}^{(1)}\right) Y_{\ell m}  = & - 2 \alpha b^3  \Bigg( f^2 Y_{\ell_1 m_1} Y_{\ell_2 m_2} Y_{\ell_3 m_3}  \phi_{\ell_1 m_1}^{+\,'}  \phi_{\ell_2 m_2}^{+\,'} \left(\phi_{\ell_3 m_3}^{+\,''} - \frac{f'}{2f}\phi_{\ell_3 m_3}^{+\,'}\right) \nonumber \\
& ~~~~~~~~~~~ + 2 \frac{f}{r^2} \gamma^{ac} Y_{\ell_2 m_2} Y_{\ell_3 m_3; a} Y_{\ell_1 m_1;c} \,\phi_{\ell_1 m_1}^+  \phi_{\ell_2 m_2}^{+\,'} \phi_{\ell_3 m_3}^{+\,'}  \nonumber \\
& ~~~~~~~~~~~ + \frac{f}{r^3} \gamma^{b d} Y_{\ell_1 m_1;c} Y_{\ell_2 m_2;d} Y_{\ell_3 m_3} \phi_{\ell_1 m_1}^+ \phi_{\ell_2 m_2}^+ \phi_{\ell_3 m_3}^{+\,'}  \nonumber \\
&~~~~~~~~~~~ + \frac{1}{r^4} \gamma^{a c} \gamma^{b d} Y_{\ell_1 m_1;c} Y_{\ell_2 m_2;d} Y_{\ell_3 m_3; b a} \,\phi_{\ell_1 m_1}^+ \phi_{\ell_2 m_2}^+ \phi_{\ell_3 m_3}^+  \Bigg)\,,
\end{align}
where~$\gamma_{ab}$ now denotes the metric on the unit 2-sphere.

Multiplying both sides with~$Y_{\ell m}^*$ and integrating over the solid angle gives
\begin{align}
{\cal L}_\ell \phi_{\ell m}^{(1)} = 
 & - 2 \alpha b^3 \Bigg\{ f^2 \phi_{\ell_1 m_1}^{+\,'}  \phi_{\ell_2 m_2}^{+\,'} \left(\phi_{\ell_3 m_3}^{+\,''} - \frac{f'}{2f}\phi_{\ell_3 m_3}^{+\,'}\right) 
\int \rd \Omega \, Y_{\ell_1 m_1} Y_{\ell_2 m_2} Y_{\ell_3 m_3}  Y_{\ell m}^* \nonumber \\
&  ~~~~~~~~~~~  + 2 \frac{f}{r^2} \phi_{\ell_1 m_1}^+ \phi_{\ell_2 m_2}^{+\,'} \phi_{\ell_3 m_3}^{+\,'}  
\int \rd \Omega \, \gamma^{ab} Y_{\ell_1 m_1;a} Y_{\ell_2 m_2} Y_{\ell_3 m_3; b} Y_{\ell m}^* \nonumber \\
&  ~~~~~~~~~~~ +  \frac{f}{r^3}  \phi_{\ell_1 m_1}^+ \phi_{\ell_2 m_2}^+ \phi_{\ell_3 m_3}^{+\,'} 
\int \rd \Omega \, \gamma^{c d} Y_{\ell_1 m_1;c} Y_{\ell_2 m_2;d} Y_{\ell_3 m_3}  Y_{\ell m}^* \nonumber \\
&  ~~~~~~~~~~~ +   \frac{1}{r^4} \phi_{\ell_1 m_1}^+ \phi_{\ell_2 m_2}^+ \phi_{\ell_3 m_3}^+  
\int \rd \Omega \, \gamma^{a c} \gamma^{b d} Y_{\ell_1 m_1;c} Y_{\ell_2 m_2;d} Y_{\ell_3 m_3; b a}  Y_{\ell m}^*\Bigg\}\,.
\end{align}
The above integrals can be collected as scalar, vector and tensor Clebsch-Gordan coefficients, defined as
\begin{align}
\mathcal{C}^{\ell m}_{\ell_1 m_1\ell_2 m_2 \ell_3 m_3} \equiv & \int \rd \Omega \, Y_{\ell_1 m_1} Y_{\ell_2 m_2} Y_{\ell_3 m_3}  Y_{\ell m}^* \,;  \nonumber \\
\mathcal{C}^{V_{13} \ell m}_{\ell_1 m_1\ell_2 m_2 \ell_3 m_3} \equiv & \int \rd \Omega \, \gamma^{ab} Y_{\ell_1 m_1;a} Y_{\ell_2 m_2} Y_{\ell_3 m_3; b} Y_{\ell m}^* \,;\nonumber \\
\mathcal{C}^{T_{13,23} \ell m}_{\ell_1 m_1\ell_2 m_2 \ell_3 m_3} \equiv & \int \rd \Omega \, \gamma^{a c} \gamma^{b d} Y_{\ell_1 m_1;c} Y_{\ell_2 m_2;d} Y_{\ell_3 m_3; b a}  Y_{\ell m}^* \,.
\label{CVT}
\end{align}
Thus the equation of motion becomes 
\begin{align}
{\cal L}_\ell \phi_{\ell m}^{(1)} &=  - 2 \alpha b^3 \lp f^2 \mathcal{C}^{\ell m}_{\ell_1 m_1\ell_2 m_2 \ell_3 m_3} \,\phi_{\ell_1 m_1}^{+\,'}  \left(\phi_{\ell_3 m_3}^{+\,''} + \frac{f'}{2f}\phi_{\ell_3 m_3}^{+\,'}\right) + 2 \frac{f}{r^2}\, \mathcal{C}^{V_{13} \ell m}_{\ell_1 m_1\ell_2 m_2 \ell_3 m_3}\, \phi_{\ell_1 m_1}^+ \phi_{\ell_3 m_3}^{+\,'} \rp \phi_{\ell_2 m_2}^{+\,'} \nonumber \\
&  ~~~ - 2 \alpha b^3\Bigg(\frac{f}{r^3} \mathcal{C}^{V_{12} \ell m}_{\ell_1 m_1,\ell_2 m_2, \ell_3 m_3}\phi_{\ell_1 m_1}  \phi_{\ell_3 m_3}^{+\,'}  + \frac{1}{r^4}\, \mathcal{C}^{T_{13,23} \ell m}_{\ell_1 m_1\ell_2 m_2 \ell_3 m_3}\, \phi_{\ell_1 m_1}^+  \phi_{\ell_3 m_3}^+ \Bigg) \phi_{\ell_2 m_2}^+ \,.
\end{align}
This matches Eq.~\eqref{Box phi 1 dphi4} in the main text.

\section{Simplification of the source term for the non-linear sigma model}
\label{source term sigma model}

In this Appendix we provide the computation for the source term in Eqs.~\eqref{NL sig example 1st} at first order, and in Eqs.~\eqref{NL sig example 2nd a}-\eqref{NL sig example 2nd b} at second order, for the example case described for the non-linear sigma model. 

\subsection{First-order equations}

The equations of motion for the fields at first order in~$\alpha$ and~$\beta$ are given in Eqs.~\eqref{EOM sigma 1st}
as
\begin{align}
\Box \phi^{(1)} & = - \alpha \phi^+ \big(\partial \chi^{(0)}\big)^2 - \beta \big(\partial \phi^{(0)}\big)^2\,; \nonumber \\
\Box \chi^{(1)} & = - \alpha \chi^{(0)} \big(\partial \phi^{(0)}\big)^2\,.
\end{align}
Expanding in spherical harmonics and specializing to the Schwarzschild background, the equations become
\begin{align}
\left({\cal L}_\ell \phi_{\ell m}^{(1)}\right) Y_{\ell m}  & =  - \alpha b^3 \phi_{\ell_3 m_3}^{+} Y_{\ell_3 m_3} \lp  f \phi_{\ell_1 m_1}^{+ \, '} \phi_{\ell_2 m_2}^{+ \, '} Y_{\ell_1 m_1} Y_{\ell_2 m_2} + \frac{1}{r^2} \gamma^{ab} \phi_{\ell_1 m_1}^{+} \phi_{\ell_2 m_2}^{+} Y_{\ell_1 m_1; a} Y_{\ell_2 m_2; b} \rp \nonumber \\
& ~~~\, - \beta b^2 \lp  f \phi_{\ell_1 m_1}^{+ \, '} \phi_{\ell_2 m_2}^{+ \, '} Y_{\ell_1 m_1} Y_{\ell_2 m_2} + \frac{1}{r^2} \gamma^{ab} \phi_{\ell_1 m_1}^+ \phi_{\ell_2 m_2}^+ Y_{\ell_1 m_1; a} Y_{\ell_2 m_2; b} \rp\,;
\nonumber \\
\left({\cal L}_\ell\chi_{\ell m}^{(1)}\right) Y_{\ell m}  & = - \alpha b^3 \phi_{\ell_3 m_3}^{+} Y_{\ell_3 m_3} \lp  f \phi_{\ell_1 m_1}^{+ \, '} \phi_{\ell_2 m_2}^{+ \, '} Y_{\ell_1 m_1} Y_{\ell_2 m_2} + \frac{1}{r^2} \gamma^{ab} \phi_{\ell_1 m_1}^+ \phi_{\ell_2 m_2}^+ Y_{\ell_1 m_1; a} Y_{\ell_2 m_2; b} \rp\,.
\end{align}
Multiplying both sides with~$Y_{\ell m}^*$ and integrating over the solid angle gives
\begin{align}
{\cal L}_\ell \phi_{\ell m}^{(1)}  & =  - \alpha b^3  \phi_{\ell_3 m_3}^+ \Bigg(  f \phi_{\ell_1 m_1}^{+ \, '} \phi_{\ell_2 m_2}^{+ \, '} \int \rd \Omega\, Y_{\ell_1 m_1} Y_{\ell_2 m_2} Y_{\ell_3 m_3} Y_{\ell m}^*   \nonumber \\ 
& ~~~~~~~~~~~~~~~~~~~  + \frac{1}{r^2} \phi_{\ell_1 m_1}^+ \phi_{\ell_2 m_2}^+ \int \rd \Omega \, \gamma^{ab} Y_{\ell_1 m_1; a} Y_{\ell_2 m_2; b} Y_{\ell_3 m_3} Y_{\ell m}^* \Bigg) \nonumber \\
& ~~~ - \beta b^2 \lp  f \phi_{\ell_1 m_1}^{+\, '} \phi_{\ell_2 m_2}^{+\, '} \int \rd \Omega \, Y_{\ell_1 m_1} Y_{\ell_2 m_2} Y_{\ell m}^* + \frac{1}{r^2} \phi_{\ell_1 m_1}^+ \phi_{\ell_2 m_2}^+  \int \rd \Omega\,  \gamma^{ab} Y_{\ell_1 m_1; a} Y_{\ell_2 m_2; b} Y_{\ell m}^* \rp \,,
\nonumber \\
{\cal L}_\ell \chi_{\ell m}^{(1)}  & =  - \alpha b^3  \phi_{\ell_3 m_3}^+ \Bigg(  f \phi_{\ell_1 m_1}^{+ \, '} \phi_{\ell_2 m_2}^{+ \, '} \int \rd \Omega \, Y_{\ell_1 m_1} Y_{\ell_2 m_2} Y_{\ell_3 m_3} Y_{\ell m}^*   \nonumber \\ 
& ~~~~~~~~~~~~~~~~~~~  + \frac{1}{r^2} \phi_{\ell_1 m_1}^+ \phi_{\ell_2 m_2}^+ \int \rd \Omega\,  \gamma^{ab} Y_{\ell_1 m_1; a} Y_{\ell_2 m_2; b} Y_{\ell_3 m_3} Y_{\ell m}^* \Bigg)\,.
\end{align}
By introducing scalar and vector Clebsch-Gordan coefficients, the equations take the simplified form given by Eqs.~\eqref{NL sig example 1st}.

\subsection{Second-order equations}

The equations of motion at second order in perturbation theory, given by Eqs.~\eqref{EOM sigma 2nd}, are reproduced here for convenience:
\begin{align}
\Box \phi^{(2)} & =  \alpha^2 \phi^{(0)} \chi^{(0) \,2} \big(\partial \phi^{(0)}\big)^2 - \alpha \phi^{(1)} \big(\partial \chi^{(0)}\big)^2 
- 2\alpha \phi^{(0)} \partial^\mu \chi^{(0)} \partial_\mu \chi^{(1)} \nonumber \\
& ~~~ + 2 \alpha \beta \phi^{(0)\, 2} \big(\partial \chi^{(0)}\big)^2 + 2 \beta^2 \phi^{(0)} \big(\partial \phi^{(0)}\big)^2 - 2 \beta \partial^\mu \phi^{(0)}\partial_\mu \phi^{(1)}\,, \nonumber \\
\Box \chi^{(2)} & =  \alpha^2 \chi^{(0)} \phi^{(0)\, 2} \big(\partial \chi^{(0)}\big)^2  - \alpha \chi^{(1)} \big(\partial \phi^{(0)}\big)^2 
- 2\alpha \chi^{(0)} \partial^\mu \phi^{(0)} \partial_\mu \phi^{(1)} \nonumber \\
& ~~~ + \alpha \beta \phi^{(0)} \chi^{(0)} \big(\partial \phi^{(0)}\big)^2\,.
\end{align}
Following similar steps as before, the~$\phi$ equation of motion becomes
\begin{align}
{\cal L}_\ell \phi^{(2)}_{\ell m}  & = \alpha^2 b^5 \phi^+_{\ell_5 m_5} \phi^+_{\ell_4 m_4} \phi^+_{\ell_3 m_3} \Bigg(f \phi_{\ell_1 m_1}^{+\,'} \phi_{\ell_2 m_2}^{+\,'} \int \rd \Omega \, Y_{\ell_1 m_1} Y_{\ell_2 m_2}
Y_{\ell_3 m_3} Y_{\ell_4 m_4} Y_{\ell_5 m_5} Y_{\ell m}^*
\nonumber \\
& ~~~~~~~~~~~~~~~~~~~~~~~~~~~~~~~~~ + \frac{1}{r^2} \phi_{\ell_1 m_1}^{+} \phi_{\ell_2 m_2}^{+} \int \rd \Omega \,\gamma^{ab}  Y_{\ell_1 m_1; a} Y_{\ell_2 m_2; b} Y_{\ell_3 m_3} Y_{\ell_4 m_4} Y_{\ell_5 m_5} Y_{\ell m}^* \Bigg)\nonumber \\
& - \alpha b^3  \phi^{(1)}_{\ell_3 m_3} \Bigg(f\phi_{\ell_1 m_1}^{+\, '} \phi_{\ell_2 m_2}^{+\, '} \int \rd \Omega \, Y_{\ell_1 m_1} Y_{\ell_2 m_2}
Y_{\ell_3 m_3} Y_{\ell m}^* 
\nonumber \\
& ~~~~~~~~~~~~~~~~~ + \frac{1}{r^2} \phi_{\ell_1 m_1}^{+} \phi_{\ell_2 m_2}^{+}  \int \rd \Omega \, \gamma^{ab}  Y_{\ell_1 m_1; a} Y_{\ell_2 m_2; b} Y_{\ell_3 m_3}  Y_{\ell m}^*\Bigg)
\nonumber \\
& - 2\alpha b^3 \phi^{+}_{\ell_3 m_3} \Bigg( f \chi_{\ell_1 m_1}^{(1)\, '} \phi_{\ell_2 m_2}^{+\,'} \int \rd \Omega \, Y_{\ell_1 m_1} Y_{\ell_2 m_2}
Y_{\ell_3 m_3} Y_{\ell m}^*
\nonumber \\
& ~~~~~~~~~~~~~~~~~ +\frac{1}{r^2} \chi_{\ell_1 m_1}^{(1)} \phi_{\ell_2 m_2}^{+}  \int \rd \Omega\, \gamma^{ab}  Y_{\ell_1 m_1; a} Y_{\ell_2 m_2; b} Y_{\ell_3 m_3}  Y_{\ell m}^*\Bigg)\,.
\end{align}
Expressing the angular integrals as scalar and vector Clebsch-Gordan coefficients, one obtains Eq.~\eqref{NL sig example 2nd a}. The derivation of Eq.~\eqref{NL sig example 2nd b} for~$\chi^{(2)}_{\ell m}$ proceeds almost identically.


\end{appendices}
 
\bibliographystyle{JHEP}
\bibliography{draft.bib}
\end{document}